\documentclass[final,5p,times,twocolumn,authoryear]{elsarticle}

\usepackage{amssymb}
\usepackage{lipsum}
\usepackage{amsmath}	
\usepackage{array}
\usepackage{multirow}
\usepackage[colorlinks=true,linkcolor=blue,urlcolor=blue]{hyperref}
\usepackage{aas_macros}
\usepackage[utf8]{inputenc} 

\journal{Astronomy $\&$ Computing}
\DeclareUnicodeCharacter{2212}{-}

\begin{document}

\begin{frontmatter}

\title{Constraining Galaxy-Halo Connection Using Machine Learning}

\author[1]{Abhishek Jana\corref{cor1}}
\ead{ajana@ksu.edu}

\author[1,2,3]{Lado Samushia}
\ead{lado@ksu.edu}

\cortext[cor1]{Corresponding author}

\affiliation[1]{organization={Department of Physics, Kansas State University},
            addressline={116 Cardwell Hall}, 
            city={Manhattan},
            postcode={66502}, 
            state={KS},
            country={USA}}

\affiliation[2]{organization={E.Kharadze Georgian National Astrophysical Observatory},
            addressline={47/57 Kostava St.}, 
            city={Tbilisi},
            postcode={0179}, 
            country={Georgia}}
            
\affiliation[3]{organization={School of Natural Sciences and Medicine, Ilia State University},
            addressline={3/5 Cholokashvili Ave.}, 
            city={Tbilisi},
            postcode={0162}, 
            country={Georgia}}

\begin{abstract}
We investigate the potential of machine learning (ML) methods to model small-scale galaxy clustering for constraining Halo Occupation Distribution (HOD) parameters. Our analysis reveals that while many ML algorithms report good statistical fits, they often yield likelihood contours that are significantly biased in both mean values and variances relative to the true model parameters. This highlights the importance of careful data processing and algorithm selection in ML applications for galaxy clustering, as even seemingly robust methods can lead to biased results if not applied correctly. ML tools offer a promising approach to exploring the HOD parameter space with significantly reduced computational costs compared to traditional brute-force methods if their robustness is established. Using our ANN-based pipeline, we successfully recreate some standard results from recent literature. Properly restricting the HOD parameter space, transforming the training data, and carefully selecting ML algorithms are essential for achieving unbiased and robust predictions. Among the methods tested, artificial neural networks (ANNs) outperform random forests (RF) and ridge regression in predicting clustering statistics, when the HOD prior space is appropriately restricted. We demonstrate these findings using the projected two-point correlation function ($w_\mathrm{p}(r_\mathrm{p})$), angular multipoles of the correlation function ($\xi_\ell(r)$), and the void probability function (VPF) of Luminous Red Galaxies from Dark Energy Spectroscopic Instrument mocks. Our results show that while combining $w_\mathrm{p}(r_\mathrm{p})$ and VPF improves parameter constraints, adding the multipoles $\xi_0$, $\xi_2$, and $\xi_4$ to $w_\mathrm{p}(r_\mathrm{p})$ does not significantly improve the constraints. 

\end{abstract}




\end{frontmatter}




\section{Introduction}
\label{introduction}


Clustering of matter on small scales due to gravitational instability is a
highly non-linear process that can only be solved by running time-consuming
N-body simulations.  The N-body simulations trace the evolution of a large
number of dark matter particles over cosmic times. They can be used to study the
gravitational collapse of dark matter particles into virialized halos. Multiple
high-performance N-body simulation codes are available that allow us to study
the clustering properties of halos, the distribution of their properties, and
the distribution of dark matter inside them, with a sub-percent precision
\citep{2001NewA....6...79S,2002A&A...385..337T,2005MNRAS.364.1105S,2009PASJ...61.1319I,2010MNRAS.401..791S,2012ASPC..453..325B,2012MNRAS.426.2046A,2015MNRAS.447.1319F,2016MNRAS.457.4340K,2016MNRAS.463.3948D,2017ComAC...4....2P,2019ApJ...875...69D,2019ApJS..245...16H,2021MNRAS.506.2871S,2021MNRAS.508.4017M}.

The creation of galaxies inside dark matter halos is a more complicated process, requiring us to keep track of multiple intricate baryonic processes occurring during galaxy formation and evolution. Even though several codes for simulating these baryonic processes are available, they are much more computationally demanding \citep{2014MNRAS.444.1453D,2015MNRAS.446..521S,2015A&C....13...12N,2017MNRAS.465.2936M,2019MNRAS.486.2827D,2019ComAC...6....2N}.
The best current pure dark matter N-body simulations have cumulative volumes of
up to $(11.19\ \mathrm{Gpc})^3$, while the best simulations with baryons in them have only $(740\ \mathrm{Mpc})^3$.

One way to circumvent this problem is to populate the N-body simulations with
galaxies based on a simplified prescription. One such approach is the Halo
Occupation Distribution (HOD) \citep{1998ApJ...494....1J,2000MNRAS.318..203S,2000MNRAS.318.1144P,2001ApJ...546...20S,2002ApJ...575..587B,2002PhR...372....1C,2005ApJ...633..791Z,2007ApJ...667..760Z,2009ApJ...707..554Z}. The main assumption of the HOD approach is that the average properties
of galaxies depend on their local environment. In the simplest version of the
HOD, we put galaxies into dark matter halos based on the host halo mass. The
host halo mass determines the probability of that halo hosting a central galaxy
and a certain number of satellite galaxies. The velocities and positions of
those galaxies are similarly determined. The exact dependence of these
probabilities on the halo mass is tuned to match the observed clustering of
galaxies. More advanced HOD prescriptions will also make the probabilities
depend on other properties of the local neighborhood, like the local density of
halos. These extra dependencies, collectively known as the assembly bias, have
been shown to be important in the simulations~\citep{2018ApJ...853...84Z,2018MNRAS.480.3978A,2019MNRAS.490.5693B,2020A&A...638A..60A,2021A&A...654A..62A}, but have not been convincingly detected in real data~\citep{2022A&A...665A..44A}.

To derive cosmological constraints from small-scale clustering we need to
marginalize all possible HOD parameters that are consistent with the data at a
given cosmology. This process could be very time-consuming. The space of HOD
parameters is at least five dimensional and the statistics used to describe
small-scale clustering are time-consuming to compute. One way to speed up the
process that is becoming popular is to use machine learning to predict
clustering as a function of HOD parameters.

The use of machine learning (ML) methods to predict cosmological observables is
currently a very popular direction of research. The artificial neural networks (ANN) have been used in
\citet{2018MNRAS.478.3410A, 2021arXiv210400595G,2023JCAP...05..025N,2023ApJ...954...11P} to efficiently model galaxy-halo properties, emulate cosmological calculations, and constrain cosmological parameters. Recent works on deriving HOD constraints using ML methods
mostly focus on methods like Symbolic Regression, Random Forest Regression,
k-nearest-neighbors \citep{Xu:2013bka,2020ApJ...889..151N,2020arXiv201200111W,2021MNRAS.507.4879X,Delgado:2021cuw,2023arXiv231110425D}.

In this work, we will train
ANN so that they can predict the values of key
clustering measures as functions of the HOD parameters. This enables us to run
Markov Chain Monte Carlo (MCMC) over the HOD parameters that converge in a
few minutes, whereas a brute force approach would take several days. We also investigate the potential of other ML techniques
like random forests (RF), and Bayesian ridge (BR) to accelerate HOD
modeling. We focus on modeling two well-known clustering metrics -- the projected
two-point correlation function $w_\mathrm{p}(r_\mathrm{p})$ and the void probability function VPF measured from mock galaxy catalogs. We also measured monopole ($\xi_0$), quadrupole ($\xi_2$), and heaxdecapole ($\xi_4$) to compare with $w_\mathrm{p}(r_\mathrm{p})$. Our goal is to train ML algorithms to rapidly predict these clustering signals for any point in HOD parameter space.

\section{Creating Mock Galaxy Catalogs}
\label{sec:hod}

To create mock galaxy catalogs from N-body simulations we perform steps described in the following subsections.

\subsection{Simulation Details}

For our analysis, we use the  {\fontfamily{qcr}\selectfont AbacusSummit\_base\_c132\_ph000} broader emulator grid of {\fontfamily{qcr}\selectfont AbacusSummit}\footnote{\url{https://abacussummit.readthedocs.io/en/latest/}} N-body simulation suite \citep{2009PhDT.......175M,2019MNRAS.485.3370G,2021MNRAS.508..575G,2018ApJS..236...43G,Garrison:2016vvp}. The simulation is conducted in a flat $\Lambda$CDM cosmology characterized by the following parameters: $\omega_\mathrm{b} = 0.02237$, $\omega_\mathrm{cdm} = 0.1200$, $h = 0.6736$, $A_\mathrm{s} = 2.1791\times10^{-9}$, and $n_\mathrm{s} = 0.9049$. The simulation box has a side length of 2000$h^{-1}\ \mathrm{Mpc}$ and contains $6912^3$ particles, totaling approximately 330 billion particles, with a particle mass of about $2\times10^9 M\textsubscript{\(\odot\)}/h$.

Our study specifically focuses on box number 132 within this simulation suite, corresponding to a redshift of $z = 1.1$. We populate this box with galaxies using the HOD model, with parameters listed in Table~\ref{tab:par}. These galaxies are used for subsequent clustering analyses as part of our broader study of large-scale structures.

\subsection{HOD Model}

We follow the HOD approach to create galaxy catalogs out of the AbacusSummit
simulation.  Given a halo mass ($M_\mathrm{halo}$), this model specifies the
probability of central ($N_\mathrm{cen}$) and satellite ($N_\mathrm{sat}$)
galaxies. We will use a simple implementation of this idea where the
probabilities only depend on the halo mass ($M_\mathrm{halo}$).  We consider central and satellite separately as central galaxies reside in the potential well of the host halo and satellite galaxies follow different physics for their formation. 
The mean occupation function of the central galaxy 
$\langle{N_\mathrm{cen}(M_\mathrm{halo})}\rangle$ is a step-like function with a
characteristic minimum halo mass $M_\mathrm{cut}$, and the mean occupation
function of the central galaxy $\langle{N_\mathrm{sat}(M_\mathrm{halo})}\rangle$
follows a power law with index $\alpha$ and characteristic satellite mass $M_1$.
For Luminous Red Galaxies (LRG), the HOD model is defined as a five-parameter
model \citep{Alam:2019pwr} with the average number density
$\langle{N(M_\mathrm{halo})}\rangle$ of a specific halo mass $M_\mathrm{halo}$ is given by

\begin{equation}
    \langle{\text{N}(M_\mathrm{halo})}\rangle = \langle{\text{N}_\mathrm{cen}(\text{M}_\mathrm{halo})}\rangle + \langle{\text{N}_\mathrm{sat}(\text{M}_\mathrm{halo})}\rangle,
\end{equation}
where,

\begin{equation}
\label{eq:centprob}
   \langle{\text{N}_\mathrm{cen}(\text{M}_\mathrm{halo})}\rangle = \frac{1}{2}p_\mathrm{max} \text{erfc}\left(\frac{\text{log}_{10}\text{M}_\mathrm{cut} - \text{log}_{10}\text{M}_\mathrm{halo}}{\sqrt{2}\text{log}_{10}(e)\sigma_\mathrm{M}}\right), 
\end{equation}

\begin{equation}
\label{eq:satprob}
    \langle{\text{N}_\mathrm{sat}(\text{M}_\mathrm{halo})}\rangle = \mathcal{H}\left(\frac{\text{M}_\mathrm{halo}}{\kappa\text{M}_\mathrm{cut}}\right) \left(\frac{\text{M}_\mathrm{halo} - \kappa\text{M}_\mathrm{cut}}{\text{M}_1}\right)^\alpha  \langle{\text{N}_\mathrm{cen}(\text{M}_\mathrm{halo})}\rangle,
\end{equation}
where,
\begin{equation}
	\mathcal{H}\left(\frac{\text{M}_\mathrm{halo}}{\kappa\text{M}_\mathrm{cut}}\right)=
	\begin{cases}
		1, & \text{M}_\mathrm{halo} > \kappa\text{M}_\mathrm{cut}.\\
		0, & \text{otherwise.}
	\end{cases}
\end{equation}
Here $p_\mathrm{max}$ is the saturation level that controls the probability of
hosting a central galaxy. For Luminous Red Galaxy (LRG) only samples,
$p_\mathrm{max}$ is set to 1 meaning that most of the massive halos are
guaranteed to possess a central galaxy.  $M_\mathrm{cut}$ is the characteristic
minimum mass of the halos to host a central galaxy and the quantity
$\sigma_\mathrm{M}$ changes the shape of the ``central galaxy probability function'', by controlling the sharpness of the transition around $M_\mathrm{cut}$. Higher  $\sigma_\mathrm{M}$ means a smoother transition between 0 and 1 occupation probabilities.

Three additional parameters $\kappa$, $\alpha$, and $M_1$ are used to get satellite probability. $M_1$ is the characteristic satellite mass, $\alpha$ is the power law index of the satellite distribution and $\kappa$ determines the
cut-off mass in the units of $M_\mathrm{cut}$.

\subsection{Galaxy Catalogs}
\label{sec:gal_cat}

We generate 20,000 sets of HOD parameters by randomly selecting them within the bounds in 
Table~\ref{tab:par} labeled as the ``full box'', with uniform probability. We discard 
the HOD parameter sets that result in a galaxy sample with a density of less than 
$\overline{n} = 4.25\times 10^{-4}h^3\ \mathrm{Mpc}^{-3}$. At these low number densities, the clustering measurements become very noisy, making it difficult to obtain reliable interpolations. The main galaxy samples relevant for cosmological analysis from large spectroscopic surveys, either currently available or expected soon (such as the baryon oscillations spectroscopic survey \citep{2013AJ....145...10D}, dark energy spectroscopic survey \citep{2016arXiv161100036D}, Euclid satellite mission \citep{2022A&A...662A.112E}, Roman space telescope \citep{2015arXiv150303757S}) have number densities that are larger than our thresholds. Computing clustering measurements from a very large number of galaxies is time-consuming. To avoid extremely large computational times, we decided to also discard the HOD parameter sets that result in a sample with a density higher than $\overline{n} = 1.75 \times 10^{-2}h^3\ \mathrm{Mpc}^{-3}$. Both of these extreme number densities are sufficiently far away from the samples obtained in realistic galaxy surveys to be of practical importance. After imposing additional cuts on the number density we are left with 8,337 mocks out of 20,000.
 
To achieve more reliable interpolation over a smaller parameter space, we separately generate 6,000 galaxy samples within tighter HOD parameter bounds. This reduced parameter range is labeled as the ``small box'' in Table~\ref{tab:par}. The range is chosen in such a way that the galaxy number density is always well within the above-mentioned range of $4.25\times 10^{-4}h^3\ \mathrm{Mpc}^{-3} < \overline{n} < 1.75 \times 10^{-2}h^3\ \mathrm{Mpc}^{-3}$.

\begin{figure}
	\includegraphics[width=\columnwidth]{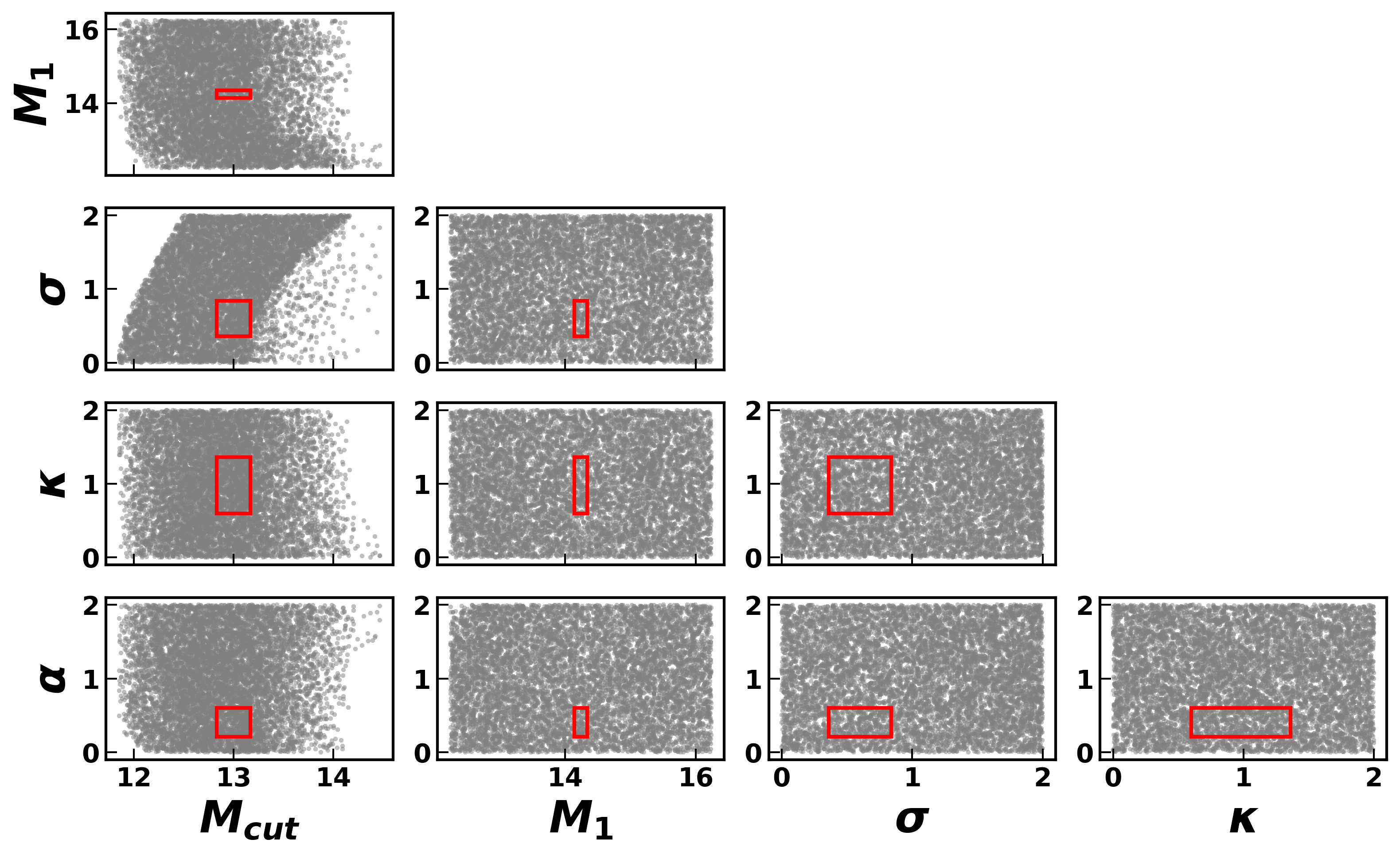}
   \caption{HOD parameter distribution of the full box after filtering. The red boxes show the small box range (see Section~\ref{sec:gal_cat}) for the range.}
   \label{fig:HOD_dist}
\end{figure}

The full box covers a very wide range of possible HOD parameter values that can describe all the galaxy samples from current and future surveys relevant to cosmological analysis. The
small box covers a range of values that is smaller than what the current precision of clustering measurements allows. We expect the interpolation
within the small box to be much more reliable. We will see in the
Section~\ref{sec:effect_of_samplesize} however even within the small box
the interpolation sometimes runs into serious problems. The size of the small box that is safe enough for interpolation will likely change for extended HOD models. Grey points on
Fig.~\ref{fig:HOD_dist} shows the HOD parameter values used in this work. The
outer edges of the plots correspond to the full box cuts, while the smaller red
rectangles correspond to the small box cuts. Some areas in the parameter space
are not covered because they were ruled out by the constraints imposed on the
number density.

For each point in the parameter space, we create a galaxy sample out of
AbacusSummit simulation. We put a central galaxy in the center of mass of the
dark matter halo based on a probability computed with Equation~\ref{eq:centprob}.
We assign the mean velocity of halo particles to this galaxy. We decide how many
satellite galaxies to put in each halo by randomly drawing a number from a
Poisson distribution with the mean given by Equation~\ref{eq:satprob}. We then
randomly pick that many dark matter particles from the halo and assign their
position and velocity to the satellites.

\section{Measurements} 
\label{measurements}

\begin{table}
    \centering
    \begin{tabular}{|p{\textwidth}||>
    {\centering\arraybackslash\hspace{0pt}}p{0.06\textwidth}|>
    {\centering\arraybackslash\hspace{0pt}}p{0.06\textwidth}|>
    {\centering\arraybackslash\hspace{0pt}}p{0.06\textwidth}|>
    {\centering\arraybackslash\hspace{0pt}}p{0.06\textwidth}|>
    {\centering\arraybackslash\hspace{0pt}}p{0.05\textwidth}|}
    \hline
    \multicolumn{1}{|c|}{\multirow{2}{*}{Parameters}} & \multicolumn{1}{l|}{\multirow{2}{*}{Fiducial}} & \multicolumn{2}{c|}{Full Box}  & \multicolumn{2}{c|}{Small Box} \\ \cline{3-6} 
    \multicolumn{1}{|l|}{} & \multicolumn{1}{l|}{} & \multicolumn{1}{c|}{Max} & Min & \multicolumn{1}{c|}{Max} & Min \\ \hline
    \multicolumn{1}{|c|}{log$M_\mathrm{cut}$} & 13.0  & \multicolumn{1}{c|}{14.5}  & 11.5  & \multicolumn{1}{c|}{13.17} & 12.83 \\ \hline
    \multicolumn{1}{|c|}{log$M_{1}$}   & 14.24 & \multicolumn{1}{c|}{16.24} & 12.24 & \multicolumn{1}{c|}{14.33} & 14.14 \\ \hline
    \multicolumn{1}{|c|}{$\sigma$} & 0.6 & \multicolumn{1}{c|}{2} & 0 & \multicolumn{1}{c|}{0.84}  & 0.36  \\ \hline
    \multicolumn{1}{|c|}{$\kappa$} & 0.98  & \multicolumn{1}{c|}{2} & 0  & \multicolumn{1}{c|}{1.36}  & 0.59  \\ \hline
    \multicolumn{1}{|c|}{$\alpha$}  & 0.4 & \multicolumn{1}{c|}{2}  & 0 & \multicolumn{1}{c|}{0.59}  & 0.2 \\ \hline
    \end{tabular}
    \caption{Table for the 5-parameter HOD model. Uniform samples are drawn to populate the HOD model (see Section~\ref{sec:gal_cat}).}
	\label{tab:par}
\end{table}

\begin{figure}
    \centering
	\includegraphics[width = \columnwidth]{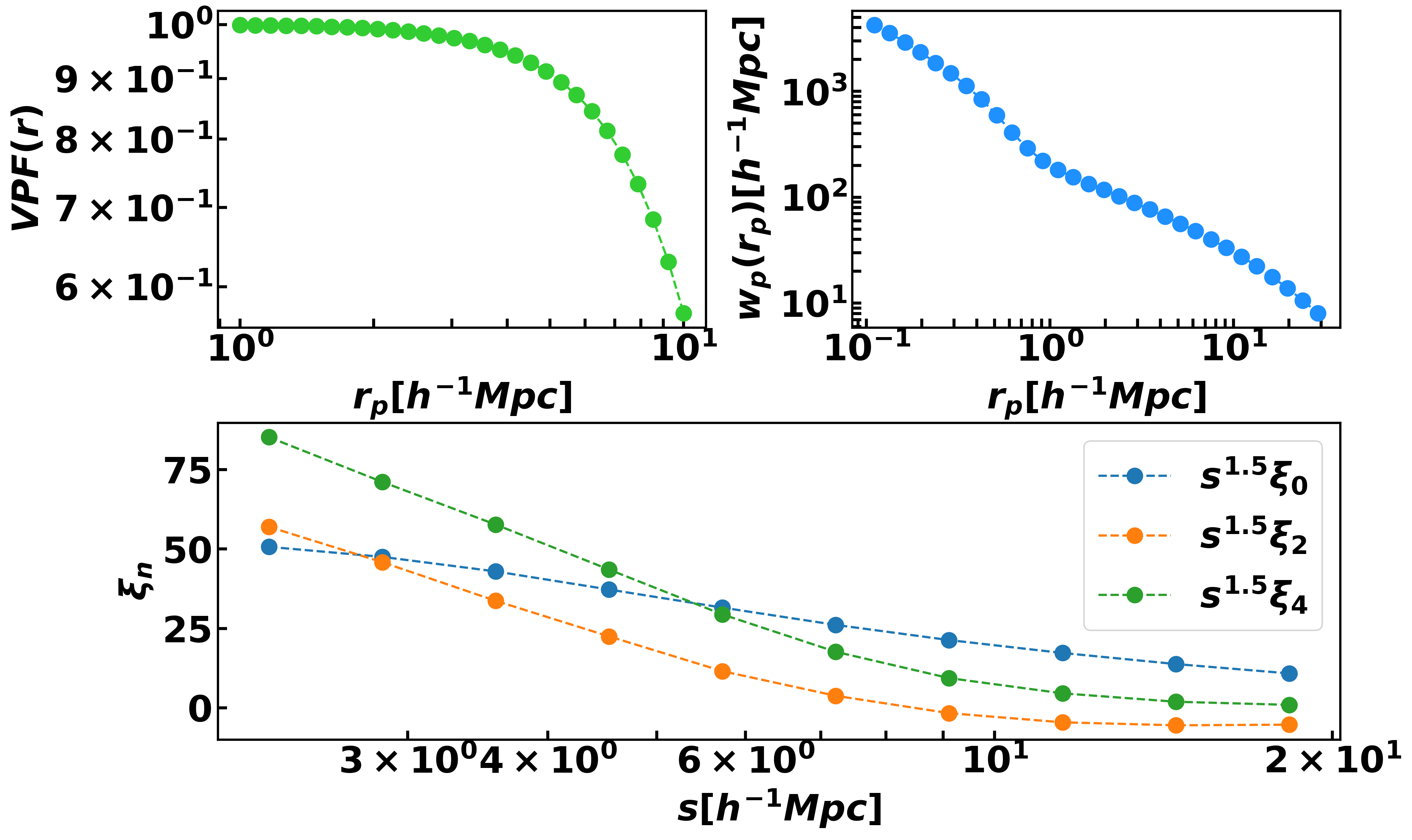}
    \caption{Small scale clustering measurements at the fiducial value from fitting the HOD model at $z = 1.1$. Left top: Void probability function as a function of $r$; Right top: projected correlation function as a function of $r_\mathrm{p}$; Bottom: correlation function as a function of $s$}
    \label{fig:stats}
\end{figure}

For each generated galaxy sample we compute a number of key clustering measures. Each measurement is computed in redshift space by transforming the galaxy coordinates from $x,y,z$ to $x,y,z+v_z(z)/aH(a)$, where we chose the z-direction to be our line of sight, $v_z$ is the z component of the galaxy velocity, $a$ is the scale factor 0.47619, and $H$ is the Hubble parameter $67.36 \mathrm{kms}^{-1}\ \mathrm{Mpc}^{-1}$.

\subsection{Projected Two-Point Correlation Function}

The main statistics that we use in our tests is the projected two-point
correlation function,

\begin{equation}
    w_\mathrm{p}(r_\mathrm{p})=2\int_{0}^{\pi_\mathrm{max}} \xi\left(r_\mathrm{p},\pi\right)d\pi.
	\label{eq:wp}
\end{equation}
Given a random galaxy, the two-point correlation function
$\xi\left(r_\mathrm{p},\pi\right)$ is the probability of finding another galaxy
within a given distance. This is a function of two variables, projected
($r_\mathrm{p}$) and  the line-of-sight ($\pi$) separation. The choice of
$\pi_\mathrm{max}$ = $60h^{-1}\ \mathrm{Mpc}$ is large enough to include most of
the correlated pairs while suppressing the noise from distant uncorrelated
pairs, giving a stable result.

We chose 30 logarithmically spaced bins ranging from $r_\mathrm{p} = 0.1 h^{-1}\
\mathrm{Mpc}$ to $r_\mathrm{p} = 31.6 h^{-1}\ \mathrm{Mpc}$.

\subsection{Void Probability Function}

To calculate the void probability function (VPF), we randomly place
$N_\mathrm{sphere}$ number of spheres of radius $r$ in the simulation box and
count the number of void spheres or the number of spheres containing no galaxy
($N_\mathrm{void}$). The ratio of the void spheres and the total number of
spheres is called the void probability function. VPF is given as

\begin{equation}
    VPF(r)=\frac{N_\mathrm{void}(r)}{N_\mathrm{sphere}(r)}.
	\label{eq:vpf}
\end{equation}
We used $N_\mathrm{sphere} = 10^{5}$. To compute VPF(r), we chose 30
logarithmically-spaced radii from r = $1h^{−1}\ \mathrm{Mpc}$ to r = $10h^{−1}\
\mathrm{Mpc}$. We use {\fontfamily{qcr}\selectfont Halotools}\footnote{\url{https://halotools.readthedocs.io/en/latest/}} (v0.7) \citep{Hearin:2016uxs} to
compute VPF and $w_\mathrm{p}$. 

\subsection{Three-Dimensional Redshift-Space Two-Point Correlation Function}
\label{sec:xi} 
We compute the auto-correlation function using the  Landy--Szalay estimator 
\citep{1993ApJ...412...64L} given by

\begin{align}
    \xi(s,\mu) &= \frac{DD(s,\mu) - 2DR(s,\mu)+RR(s,\mu)}{RR(s,\mu)},
\end{align}
where $DD$, $DR$, and $RR$ are galaxy-galaxy, galaxy-random and random-random
pair count as a function of the bin of separation vector in 3-dimensional space.
Where DD is the galaxy-galaxy pair count with certain redshift-space separation
s, and cosine of the angle of the pair to the line-of-sight $\mu$. DR is the
galaxy-random pair counts and RR is the random-random pair counts.

We then integrate over $\mu$ to obtain the angular multipoles of the correlation
function
\begin{align}
	\xi_\ell (s) &= \frac{2\ell + 1}{2} \int_{-1}^{1} \xi\left(s,\mu\right)\mathcal{L}_{\ell}(\mu) d\mu,
\end{align}
\noindent
where $\mathcal{L}_{\ell}(\mu)$ is the Legendre polynomial of order $\ell$. We
measure correlation function monopole $\xi_0(s)$, quadrupole $\xi_2$ and
hexadecapole $\xi_4$. We chose 10 bins equally spaced in $s$ between
$2h^{-1}\ \mathrm{Mpc}$ and $20h^{−1}\ \mathrm{Mpc}$ with $\mu = 100$ (see Fig.~\ref{fig:stats}).

We use {\fontfamily{qcr}\selectfont Corrfunc}
\footnote{\url{https://corrfunc.readthedocs.io/en/master/}}
\citep{2020MNRAS.491.3022S,Sinha:2019reo} for the calculation of $\xi_\ell$.

\section{Covariance}
\begin{figure}

	\includegraphics[width = \linewidth]{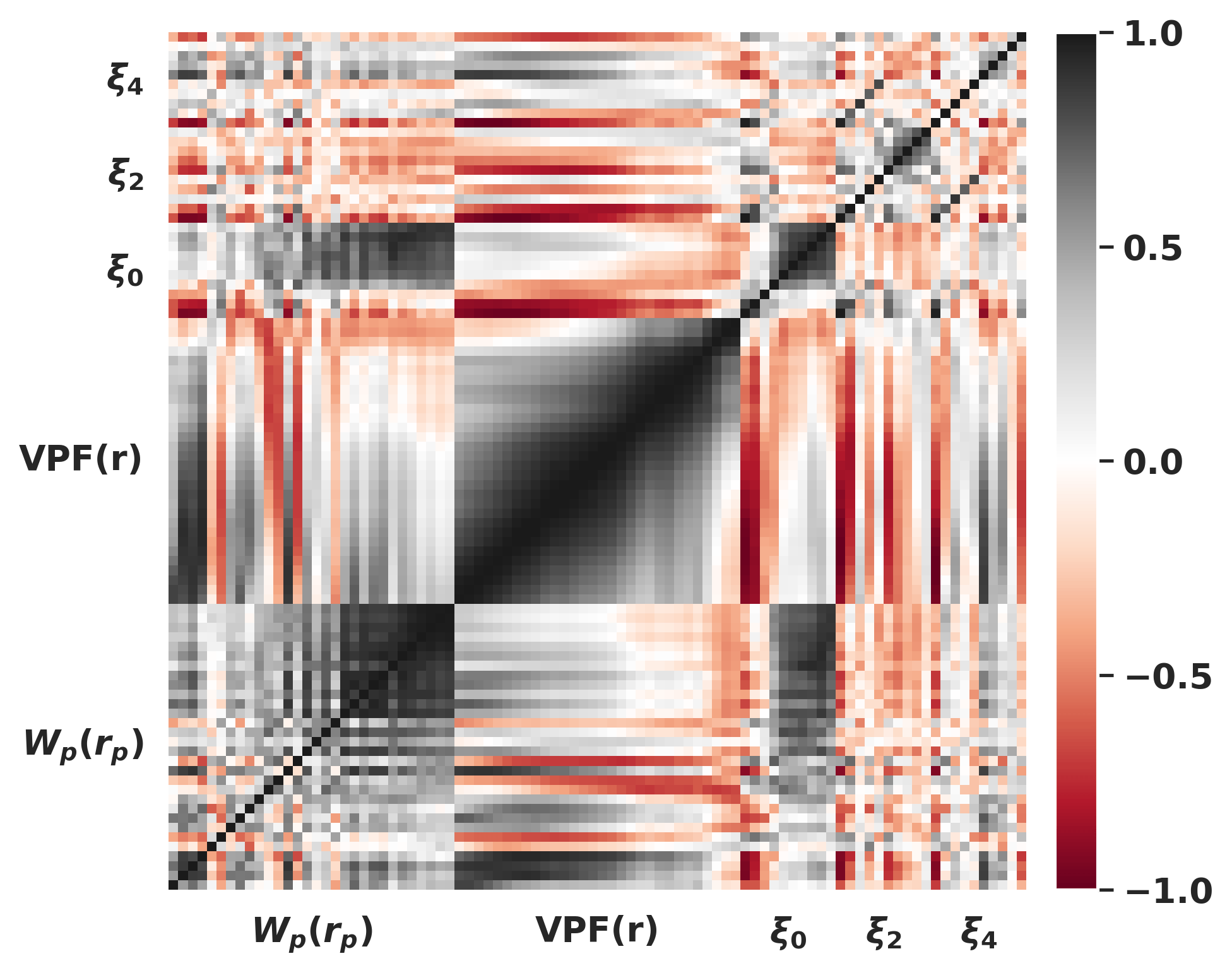} 
    \caption{Jackknife Covariance calculated at the fiducial value: from left to right 0-30 $w_\mathrm{p}(r_\mathrm{p})$, 30-60 VPF(r), 60-70 $\xi_0$, 70-80 $\xi_2$, and 80-90 $\xi_4$ }
   \label{fig:cov}
\end{figure}

We use the jackknife resampling technique
\citep{2008arXiv0805.2325M,2009MNRAS.396...19N,2016MNRAS.456.2662F,2021MNRAS.505.5833F,2022MNRAS.514.1289M}
to estimate the covariance matrix of our measurements. The simulation box is divided into 300 overlapping sub-volumes by partitioning it along the $x$, $y$, and $z$ axes in three different configurations. First, we create cuboidal sub-volumes with dimensions of $200h^{-1}\ \mathrm{Mpc}$ along the $x$ and $y$ axes and $2000h^{-1}\ \mathrm{Mpc}$ along the $z$-axis. We then repeat the process by alternating the longer $2000h^{-1}\ \mathrm{Mpc}$ axis between $y$ and $z$, ensuring that each configuration maintains the same $200h^{-1}\ \mathrm{Mpc}$ dimensions along two axes and $2000h^{-1}\ \mathrm{Mpc}$ along the third axis. This approach results in 300 overlapping cuboidal sub-volumes. We compute the clustering measurements described in Section~\ref{measurements}, each time excluding one sub-volume, generating 300 jackknife realizations.

We arrange them in the following order: The  $w_\mathrm{p}$
measurements followed by $VPF$, $\xi_\mathrm{0}$, $\xi_\mathrm{2}$, and
$\xi_\mathrm{4}$. All the measurements are arranged from the smallest scales to
the largest. The jackknife estimate of the covariance matrix is then computed
using the formula 
\begin{align}
    C_{ij} &= \frac{1}{300}\displaystyle\sum_{k,\ell=1}^{300}(X_i^k - \overline{X}_i)(X_j^\ell - \overline{X}_j),\\
    \overline{X}_i &= \frac{1}{300}\displaystyle\sum_{k=1}^{300}X_i^k,
\end{align}
where $X_i^k$ denotes the $i^\mathrm{th}$ element of the measurement vector
computed from the $k^\mathrm{th}$ jackknife realization.

The reduced covariance matrix, defined as
\begin{equation}
    R_{ij} = \frac{C_{ij}}{\sqrt{C_{ii}C_{jj}}},
\end{equation}
is shown in Fig.~\ref{fig:cov}. The $w_\mathrm{p}$ and VPF measurements between neighboring bins are strongly correlated. So are the $\xi_\ell$ measurements with different $\ell$ but similar separation. VPF is not correlated with other clustering measures. These results are in line with theoretical expectations.

These covariance matrices would not be reliable for a robust cosmological analysis. Our goal however is to study the convergence properties of the machine learning-based interpolation methods, and for our purposes, any reasonable positive-definite matrix that loosely resembles a typical covariance matrix of clustering measurements is sufficient to demonstrate the point.

\section{Machine Learning Methods}
\label{sec:method}

We used supervised ML regression for our work. A more detailed discussion on ML
methods is done in~\ref{sec:ML_overview}.  Our goal is to be able to
predict the measures of small-scale clustering as functions of HOD parameters.
Our feature columns are the 5 HOD parameters and our target columns are the
$w_\mathrm{p}$, VPF, and $\xi_\ell$ at different scales.  We go through the
following sequence of steps:

\begin{enumerate}
    \item Check the distribution of the data.
    \item Data preprocessing.
    \item Split the data into train and test.
    \item Fit different ML models to train the data.
    \item Hyperparameter tuning.
    \item Perform accuracy test.
    \item Get the best model.
\end{enumerate} 

\subsection{Check the Distribution of the Data}
\begin{figure}
    \centering
	\includegraphics[width = \columnwidth]{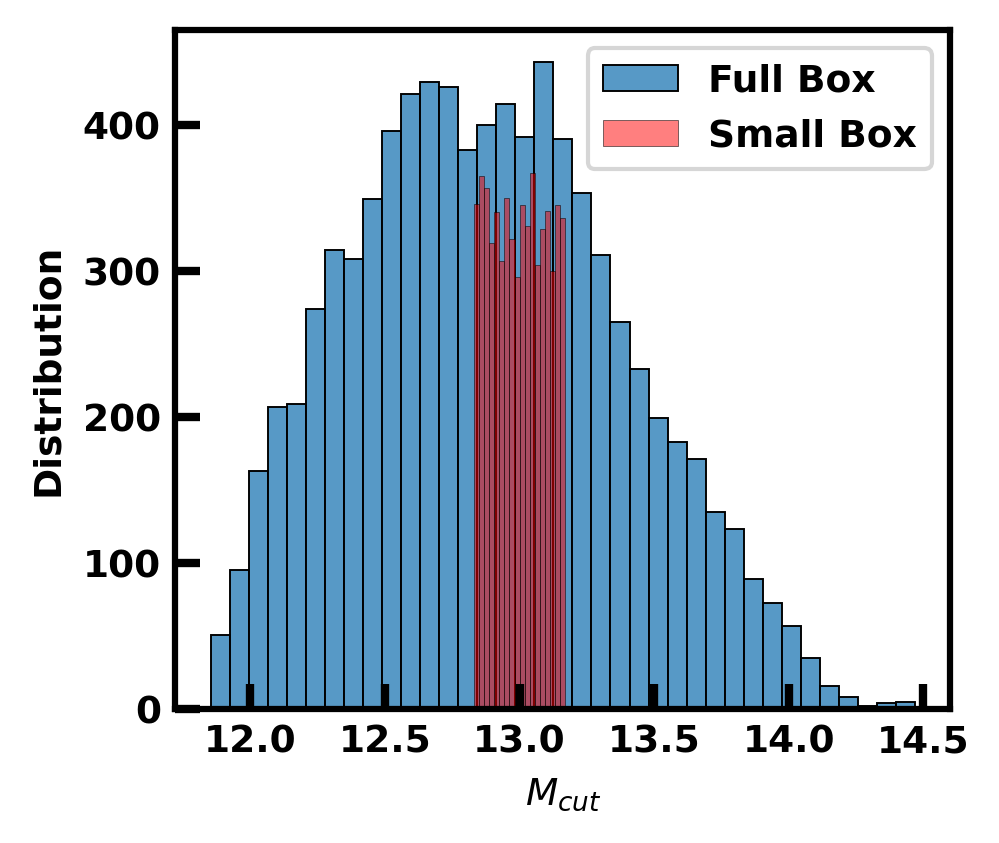}
    \caption{$M_\mathrm{cut}$ distribution on full and small boxes (see Section~\ref{sec:gal_cat}). HOD parameters are drawn uniformly for both the full box and small box and after applying filtering on the full box the distribution is nearly Gaussian. }
    \label{fig:m_cut_distribution}
\end{figure}

The distribution of the HOD parameters on both the full and small boxes is drawn
from uniform samples. However, due to the filtering method applied on the full
box, the $M_\mathrm{cut}$ parameter has a near Gaussian distribution, as shown in
Fig.~\ref{fig:m_cut_distribution}. We suspect some of the inconsistencies in the full box may arise due to this filtering (see Section~\ref{sec:prior}).

\subsection{Data Preprocessing}
We used {\fontfamily{qcr}\selectfont Scikit-Learn}
\footnote{\url{https://scikit-learn.org/stable/index.html}}
\citep{2011JMLR...12.2825P} for all the ML-related calculations.  To improve the
model's performance, we increased the number of input features by using the 
\textit{PolynomialFeatures} module followed by the \textit{StandardScaler}
module from Scikit-Learn. For the small box, we choose {\fontfamily{qcr} \selectfont PolynomialFeatures(degree=2, interaction\_only=True)}, resulting in a total of 16 input features for the small box  (Discussed in~\ref{sec:accuracy_score}). The purpose of using StandardScaler is to center the data by subtracting the mean and scaling it to
have a unit variance. This preprocessing step is crucial for standardizing the
features and ensuring that they contribute equally to the learning process,
especially when features have different scales or variances.

The target column, representing the small-scale clustering statistics, is highly
skewed which can introduce bias in model training. To avoid this issue, we
applied several common data transformation methods, including Log
transformation (LT), StandardScaler (SC), Yeo-Johnson (YJ), and Box--Cox (BC) to the raw data. These transformations were used to reduce biases, and their importance is further
discussed in Section~\ref{subsec:data_transformation}.

\subsection{Split the Data into Train and Test}

The data is split into two sets, the training set and the test set, with the
latter typically representing $20\%$ of the total data. We train the model using
the training set and evaluate its accuracy on the test set. In our study, we
used a training set of 6660 and a test set of 1666 for the full box. For the
small box, we varied the sample size to investigate its effect on the model's
performance (as shown in Section~\ref{sec:effect_of_samplesize}), with training set
sizes of 800, 1600, and 5400 and a test set size of 600.

\subsection{Fit Different ML Models to Train the Data}
For better accuracy and robustness, we evaluated several regression models in
this study. We tested the performance of three models, namely RF
regression, BR regression, and ANN. Each method is discussed
in detail in ~\ref{sec:ML_overview}.

We found out that ANN works the best among the three methods. A more detailed comparison is discussed in Section~\ref{sec:model_selection}

\begin{table}

\renewcommand{\arraystretch}{1.8} 

\centering
\begin{tabular}{| p{0.09\textwidth} ||>
    {\centering\arraybackslash\hspace{0pt}}p{0.04\textwidth}|>
    {\centering\arraybackslash\hspace{0pt}}p{0.04\textwidth}|>
    {\centering\arraybackslash\hspace{0pt}}p{0.04\textwidth}|>
    {\centering\arraybackslash\hspace{0pt}}p{0.04\textwidth}|>
    {\centering\arraybackslash\hspace{0pt}}p{0.04\textwidth}|}
\hline
\multicolumn{1}{|c|}{Statistics} & $w_\mathrm{p}$   & VPF & $\xi_0$ & $\xi_2$ & $\xi_4$ \\ \hline
lambda\_init                     & 0.01 & 0.1 & 0.1   & 0.1   & 0.1   \\ \hline
\end{tabular}
\caption{Hyperparameters of Bayesian Ridge regressor for the small box. Only lambda\_init is changed during model training keeping other hyperparameters as default.}
\label{tab:hype_br}
\end{table}

\begin{table}

\renewcommand{\arraystretch}{1.8} 

\centering
\begin{tabular}{| p{0.15\textwidth} ||>
    {\centering\arraybackslash\hspace{0pt}}p{0.037\textwidth}|>
    {\centering\arraybackslash\hspace{0pt}}p{0.037\textwidth}|>
    {\centering\arraybackslash\hspace{0pt}}p{0.037\textwidth}|>
    {\centering\arraybackslash\hspace{0pt}}p{0.037\textwidth}|>
    {\centering\arraybackslash\hspace{0pt}}p{0.037\textwidth}|}
\hline
\multicolumn{1}{|c|}{Statistics} & $w_\mathrm{p}$       & VPF      & $\xi_0$      & $\xi_2$      & $\xi_4$      \\ \hline
hidden\_layer\_sizes & 500      & 500      & 500      & 500      & 500      \\ \hline
alpha                & 0.01     & 0.01     & 0.01     & 0.01     & 0.01     \\ \hline
activation           & tanh     & tanh     & tanh     & tanh     & tanh     \\ \hline
learning\_rate\_init & $5\times10^{-4}$   & $5\times10^{-4}$   & $5\times10^{-4}$   & $5\times10^{-4}$   & $5\times10^{-4}$   \\ \hline
learning\_rate       & constant & constant & constant & constant & constant \\ \hline
solver               & adam     & adam     & adam     & adam     & adam     \\ \hline
random\_state        & 1        & 1        & 1        & 1        & 1        \\ \hline
validation\_fraction & 0.1      & 0.1      & 0.2      & 0.2      & 0.2      \\ \hline
early\_stopping      & True     & True     & True     & True     & True     \\ \hline
max\_iter            & 2000     & 5000     & 2000     & 2000     & 2000     \\ \hline
\end{tabular}
\caption{Hyperparameters of ANN for the small box.}
\label{tab:hype_ann}
\end{table}

\subsection{Hyperparameter Tuning}

Hyperparameter tuning is a crucial step in machine learning, as it helps adjust a model to prevent overfitting or underfitting. We used the Scikit-Learn library to perform the following hyperparameter tuning for three models.

For the small box dataset, we used the BayesianRidge module for the BR model. The modified hyperparameters are shown in Table~\ref{tab:hype_br}, while all other parameters were kept at their default values. We used the MLPRegressor module for the ANN model, and the corresponding hyperparameters are listed in Table~\ref{tab:hype_ann}. 

We tested the RF model only on $w_\mathrm{p}$ using the RandomForestRegressor module with {\fontfamily{qcr}\selectfont n\_estimators=200, criterion='squared\_error', bootstrap=False, random\_state=0, warm\_start=True}, while keeping all other parameters at their default settings. RF is a widely used method that has been applied to various cosmological problems in previous studies\citep{2018MNRAS.479.3405L,2018ApJ...859..129N,2020MNRAS.491.1575C,2021MNRAS.502.2770M}. RF is not a good choice for our problem as it relies on a series of decision rules based on individual features, which may not capture the full complexity of the data.


\begin{table}
\renewcommand{\arraystretch}{1.8} 
\centering
\begin{tabular}{| p{0.045\textwidth} ||>
    {\centering\arraybackslash\hspace{0pt}}p{0.05\textwidth}|>
    {\centering\arraybackslash\hspace{0pt}}p{0.04\textwidth}|>
    {\centering\arraybackslash\hspace{0pt}}p{0.04\textwidth}|>
    {\centering\arraybackslash\hspace{0pt}}p{0.03\textwidth}|>
    {\centering\arraybackslash\hspace{0pt}}p{0.03\textwidth}|}
\hline
Model  & (\%) $\overline{RMSE}$ & (\%) $\overline{MAE}$ & Adj $\overline{R^2}$  & (\%) RMSE Range & (\%) MAE Range \\ \hline
RF  & \multicolumn{1}{c|}{2.55} & \multicolumn{1}{c|}{1.57} & \multicolumn{1}{c|}{0.982} & \multicolumn{1}{c|}{4.86 - 1.26} & \multicolumn{1}{c|}{3 - 0.8} \\ \hline
BR & \multicolumn{1}{c|}{1.24} & \multicolumn{1}{c|}{0.94} & \multicolumn{1}{c|}{0.995} & \multicolumn{1}{c|}{1.78 - 0.78} & \multicolumn{1}{c|}{1.29 - 0.65} \\ \hline
ANN & \multicolumn{1}{c|}{0.63} & \multicolumn{1}{c|}{0.48} & \multicolumn{1}{c|}{0.999} & \multicolumn{1}{c|}{1.11 - 0.35} & \multicolumn{1}{c|}{0.72 - 0.27} \\ \hline
\end{tabular}
\caption{Comparison of RF, BR, ANN regression model performance on $w_\mathrm{p}$ for the small box test set. We calculate the average percent RMSE, average percent MAE, and average $R^2$ score for comparing the model performance (see Section~\ref{sec:accuracy_test} for definition).}
\label{tab:multi_output_performance}
\end{table}

\begin{table}
\centering
\begin{tabular}{| p{0.045\textwidth} ||>
    {\centering\arraybackslash\hspace{0pt}}p{0.079\textwidth}|>
    {\centering\arraybackslash\hspace{0pt}}p{0.079\textwidth}|>
    {\centering\arraybackslash\hspace{0pt}}p{0.079\textwidth}|>
    {\centering\arraybackslash\hspace{0pt}}p{0.079\textwidth}|>
    {\centering\arraybackslash\hspace{0pt}}p{0.079\textwidth}|}
\hline
Model               & Bin No. & 5 Best (\%) RMSE & Bin No. & 5 Worst (\%) RMSE \\ \hline
\multirow{5}{*}{RF} & 30                           & 1.26           & 1                            & 4.88            \\ \cline{2-5} 
                    & 29                           & 1.26           & 2                            & 4.84            \\ \cline{2-5} 
                    & 28                           & 1.28           & 3                            & 4.78            \\ \cline{2-5} 
                    & 27                           & 1.31           & 4                            & 4.72            \\ \cline{2-5} 
                    & 26                           & 134            & 5                            & 4.66            \\ \hline
\multirow{5}{*}{BR} & 30                           & 0.78           & 9                            & 1.78            \\ \cline{2-5} 
                    & 29                           & 0.8            & 6                            & 1.78            \\ \cline{2-5} 
                    & 28                           & 0.81           & 7                            & 1.77            \\ \cline{2-5} 
                    & 26                           & 0.82           & 5                            & 1.76            \\ \cline{2-5} 
                    & 27                           & 0.84           & 1                            & 1.73            \\ \hline
\multirow{5}{*}{NN} & 28                           & 0.35           & 3                            & 1.11            \\ \cline{2-5} 
                    & 29                           & 0.36           & 5                            & 1.02            \\ \cline{2-5} 
                    & 27                           & 0.36           & 1                            & 1.0             \\ \cline{2-5} 
                    & 30                           & 0.38           & 2                            & 1.0             \\ \cline{2-5} 
                    & 25                           & 0.38           & 4                            & 0.98            \\ \hline

\end{tabular}
\caption{Performance for 5 best and worst outputs of $w_\mathrm{p}$ on Box--Cox transformation for the small box test set. We compare RF, BR, and ANN models to report the percent RMSE.}
\label{tab:best_worst_outputs}
\end{table}

\begin{table}
\renewcommand{\arraystretch}{1.8} 

\centering
    \begin{tabular}{|p{0.1\textwidth}||>
    {\centering\arraybackslash\hspace{0pt}}p{0.1\textwidth}|>
    {\centering\arraybackslash\hspace{0pt}}p{0.1\textwidth}|>
    {\centering\arraybackslash\hspace{0pt}}p{0.05\textwidth}|>
    {\centering\arraybackslash\hspace{0pt}}p{0.1\textwidth}|>
    {\centering\arraybackslash\hspace{0pt}}p{0.05\textwidth}|}
    \hline
    \multirow{2}{*}{Adjusted $R^2$} & \multicolumn{3}{c|}{Small Box} & \multicolumn{2}{c|}{Full Box}  \\ \cline{2-6} & \multicolumn{1}{c|}{RF}  & \multicolumn{1}{c|}{BR}  & ANN    & \multicolumn{1}{c|}{BR}  & ANN  \\ \hline
    Unscaled  & \multicolumn{1}{c|}{0.966} & \multicolumn{1}{c|}{0.99} & 0.97 & \multicolumn{1}{c|}{0.885} & 0.98  \\ \hline
    Standard Scalar & \multicolumn{1}{c|}{0.982} & \multicolumn{1}{c|}{0.99} & 0.998 & \multicolumn{1}{c|}{0.885} & 0.994 \\ \hline
    Yeo-Jhonson  & \multicolumn{1}{c|}{0.983} & \multicolumn{1}{c|}{0.995} & 0.999 & \multicolumn{1}{c|}{--} & -- \\ \hline
    Box--Cox   & \multicolumn{1}{c|}{0.982} & \multicolumn{1}{c|}{0.995} & 0.999 & \multicolumn{1}{c|}{--} & -- \\ \hline
    \end{tabular}
    \caption{Adjusted $R^2$ score of $w_\mathrm{p}$ on Box--Cox transformation for the small box test set and full box test set. We compare RF, BR, and ANN models to report the Adjusted $R^2$ score  }
    \label{tab:r2_score}
\end{table}

\subsection{Perform Accuracy Test}
\label{sec:accuracy_test}

Numerous performance measures are employed to assess the efficacy of an ML model, such as the root mean squared error (RMSE), mean absolute error (MAE), and adjusted $R^2$ score (see ~\ref{sec:accuracy_score}). Since each bin in our clustering statistics has a different range, we utilize modified versions of RMSE and MAE namely average percent RMSE and average percent MAE respectively. 
The average percent RMSE is defined as
\begin{align}
    (\%) \overline{RMSE} &= \frac{1}{n_{bins}}\displaystyle\sum_{i=1}^{n_{bins}}\frac{RMSE_i}{\left(\frac{\left|y^i_\mathrm{max}\right| + \left|y^i_\mathrm{min}\right|}{2}\right)} \times 100.
\end{align}
The average percent MAE is defined as
\begin{align}
    (\%) \overline{MAE} &= \frac{1}{n_{bins}}\displaystyle\sum_{i=1}^{n_{bins}}\frac{MAE_i}{\left(\frac{\left|y^i_\mathrm{max}\right| + \left|y^i_\mathrm{min}\right|}{2}\right)} \times 100,
\end{align}
where $n_{bins}$ is the number of bins, and for the $i_{th}$ bin, $y^i_\mathrm{max}$ and $y^i_\mathrm{min}$ represent the maximum and minimum truth values, respectively.
Similarly, the average adjusted $R^2$ (Adj $\overline{R^2}$) score is calculated by taking the sum of total adjusted $R^2$ scores and dividing it by $n_{bins}$. We present results solely for the projected correlation function on the small box. However, we performed the same analysis on other statistics when selecting an optimal model.

In Table~\ref{tab:multi_output_performance}, we compare $(\%) \overline{RMSE}$, $(\%) \overline{MAE}$, and average adjusted $\overline{R^2}$ score on RF, BR, and ANN. The ANN outperforms RF and BR by a considerable margin. Table~\ref{tab:best_worst_outputs} displays the best 5 and worst 5 bins for $(\%) \overline{RMSE}$. All models exhibit superior performance at larger projected separations ($r_\mathrm{p}$). The results for the full box and the small box can be found in Table~\ref{tab:r2_score}. We do not include results for YJ and BC for the full box due to their inconsistency with sparse data. For all models, except BR on the full box, performance scores are acceptable (Adjusted $R^2$ score is above 0.95 on the test set). Typically, this would be considered a satisfactory performance; however, we will later demonstrate that even with such a high-performance score, the models result in strongly biased posterior likelihood contours.

Furthermore, we examine the cross-validation scores for all statistics to assess the uniformity of the data and conclude that the variance was reduced with data transformation.

\subsection{Likelihood Analysis}
We perform MCMC \citep{2013PASP..125..306F} to
obtain the posterior probability distribution of parameter space. The likelihood
function $\mathcal{L} \propto \exp(-\chi^2/2)$, where $\chi^2$ is given by

\begin{align}
	\chi^2 = (x - \mu)^T C^{-1} (x - \mu).
\end{align}
In our analysis, $x$ denotes the machine learning model predictions (`theory'), while $\mu$ represents the actual simulation measurements (`observation'). Specifically, we applied this framework to $w_\mathrm{p}$, $VPF$, and, $\xi_\ell$, as well as their combinations (see Section~\ref{sec:conclusion}). Here, $(x - \mu)$ is the difference between theory and observation and
$C^{-1}$ is the inverse of the covariance matrix.  We use {\fontfamily{qcr}\selectfont Emcee} \footnote{\url{https://emcee.readthedocs.io/en/stable/}} to estimate the posterior distribution of the model parameters. Several combinations of statistics were tested using 250 walkers, with the first 200 steps discarded as burn-in. Given the model's efficiency, we achieved convergence within a minute. Our findings are discussed in detail in the next section.

\section{Constraining HOD parameters}
\label{sec:constrain}

The ultimate objective of ML-based emulators is to be able to run MCMC chains on measured data to constrain HOD parameters. While all our approaches yield acceptable performance metrics, we will demonstrate that unless considerable care is exercised, they result in inaccurate likelihood surfaces. The cumulative effect of small inaccuracies significantly biases the best-fit values and often leads to underestimated uncertainties in derived model parameters. This holds true even when we constrain ourselves to the small box where the HOD parameter range is narrow. We will present results for the $w_\mathrm{p}$, but the general conclusions are similar for all other statistics. Despite our model's ability to accurately predict the full projected correlation function, we only selected $r_\mathrm{p} = 1.77 h^{-1}\ \mathrm{Mpc}$ to $r_\mathrm{p} = 17.78 h^{-1}\ \mathrm{Mpc}$ (bin 15 -- bin 27) for our MCMC run. This was done to maintain consistency with the $\xi_\ell$ range (s = $2h^{-1}\ \mathrm{Mpc} - 20h^{-1}\ \mathrm{Mpc}$), as we will demonstrate how HOD constraints improve with the combination of clustering statistics (see Section~\ref{sec:conclusion}).

\subsection{Effect of Data Transformations}
\label{subsec:data_transformation}

\begin{figure*}
    \centering
	\includegraphics[width = \linewidth]{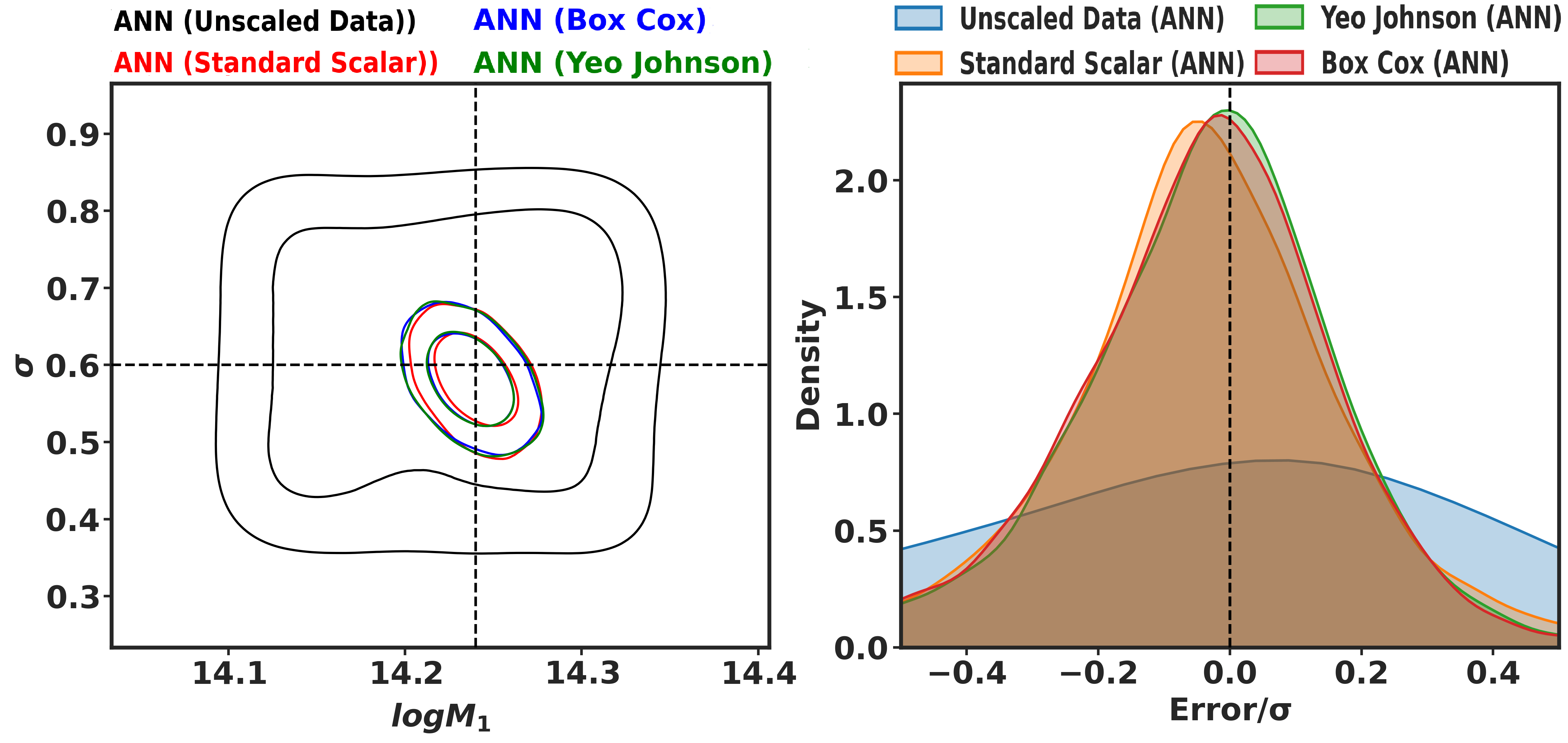}
    \caption{Exploring the impact of data transformation on ANN Method within the small Box. Left: Equipotential contours of the MCMC posterior from fitting the HOD model at $z=1.1$ using $w_\mathrm{p}$. The plot shows the convergence between transformed and non-transformed data. Right: Density plot between Transformed vs. Non-transformed Data as a function of Error/$\sigma$. The $\sigma$ is calculated from the jackknife covariance matrix.  }
    \label{fig:comparison_transformed_not_transformed_ANN}
\end{figure*}

We find that data transformations are essential. Without them, the likelihood
contours are strongly biased. Fig.~\ref{fig:comparison_transformed_not_transformed_ANN} shows the distribution of the inaccuracy in the ANN-based emulator prediction at different HOD
parameter values (left panel), and the resulting one and two $\sigma$ contours in the $\sigma$ vs $logM_1$ plane. We compute this for unscaled data and three different data transformation algorithms: SC, BC, and, YJ. The likelihood contours are strongly biased when no transformation is applied. However, applying a data transformation corrects this bias. 

Despite the standard scalar transformation exhibiting an off-center error distribution, we find that all three approaches perform equally well when it comes to the likelihood contours. The bias in the computed likelihood function is much larger than what the left panel of Fig.~\ref{fig:comparison_transformed_not_transformed_ANN} would suggest. The error in the density plots, represented as $Error/\sigma$, is calculated as the difference between the predicted values from the machine learning models and the true values obtained from the simulation, normalized by the standard deviation ($\sigma$) derived from the jackknife covariance matrix. Specifically, for each predicted value $y_i$, the error is computed as $Error_i = y_i^\mathrm{pred} - y_i^{\mathrm{true}}$, where $\sigma_i$ is the jackknife estimate of the standard deviation for the $i^{th}$ data point. This normalization allows for a clearer interpretation of the prediction errors by accounting for the intrinsic variability in the data. All the Error calculations shown in this work are calculated on $w_\mathrm{p}$. However, we found similar results for the other measurements ($VPF$, $\xi_\ell$) as well.

\begin{figure*}
    \centering
	\includegraphics[width = \linewidth]{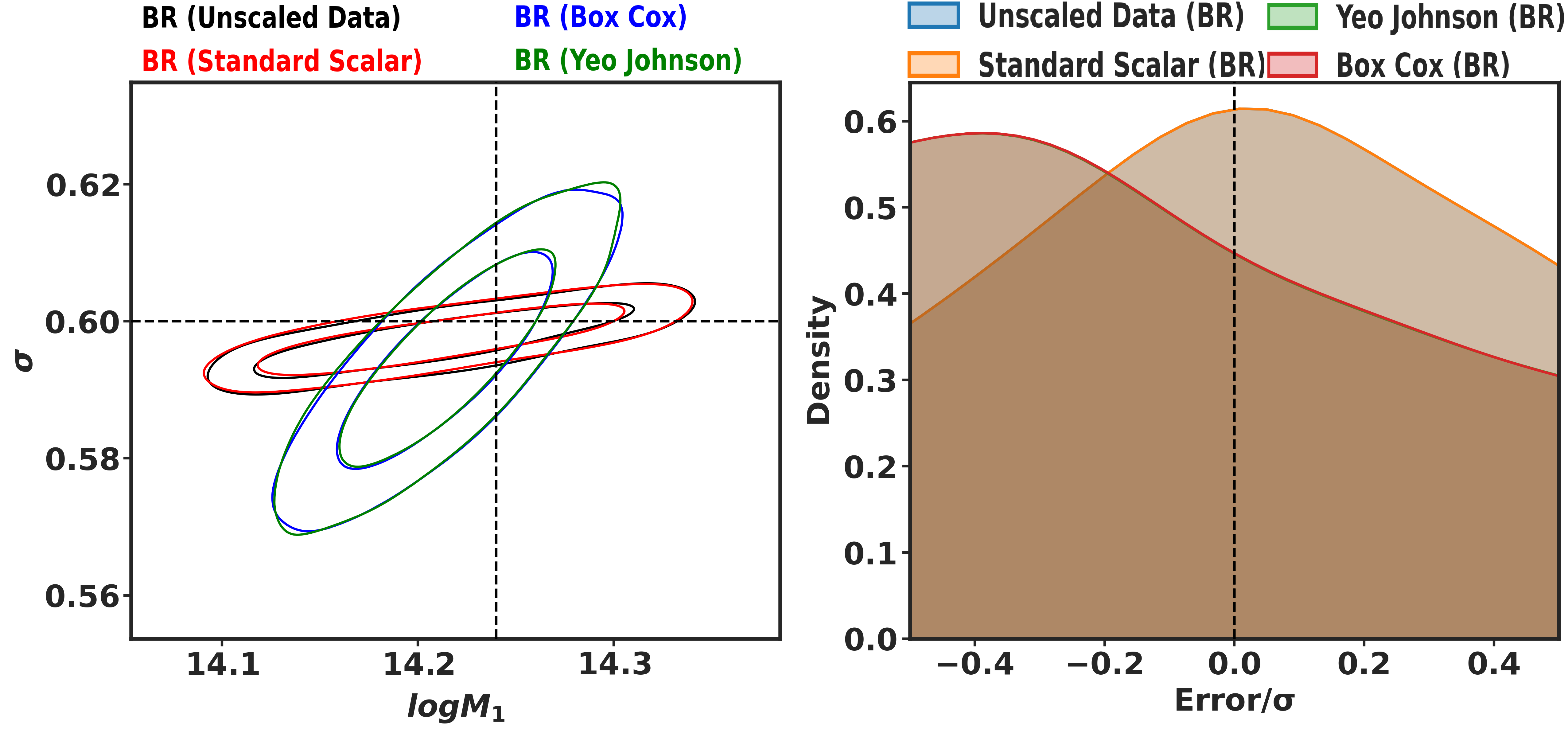}
    \caption{Exploring the impact of data transformation on BR Method within the small Box. Left: Equipotential contours of the MCMC posterior from fitting the HOD model at $z=1.1$ using $w_\mathrm{p}$. The plot shows the convergence between transformed and non-transformed data. Right: Density plot between Transformed vs. Non-transformed Data as a function of Error/$\sigma$. The $\sigma$ is calculated from the jackknife covariance matrix.   }
    \label{fig:comparison_transformed_not_transformed_BR}
\end{figure*}

Fig.~\ref{fig:comparison_transformed_not_transformed_BR} shows similar
computation for the BR method. The data scaling is essential here as well.
Unlike the ANN, however, none of the transformations can produce
unbiased likelihood contours, even though the YJ and the BC transformations lead
to consistent results. Here as well, the bias in the likelihood contours is much
more visible than what the error distribution plot would suggest.


\begin{figure*}
    \centering
	\includegraphics[width = \linewidth]{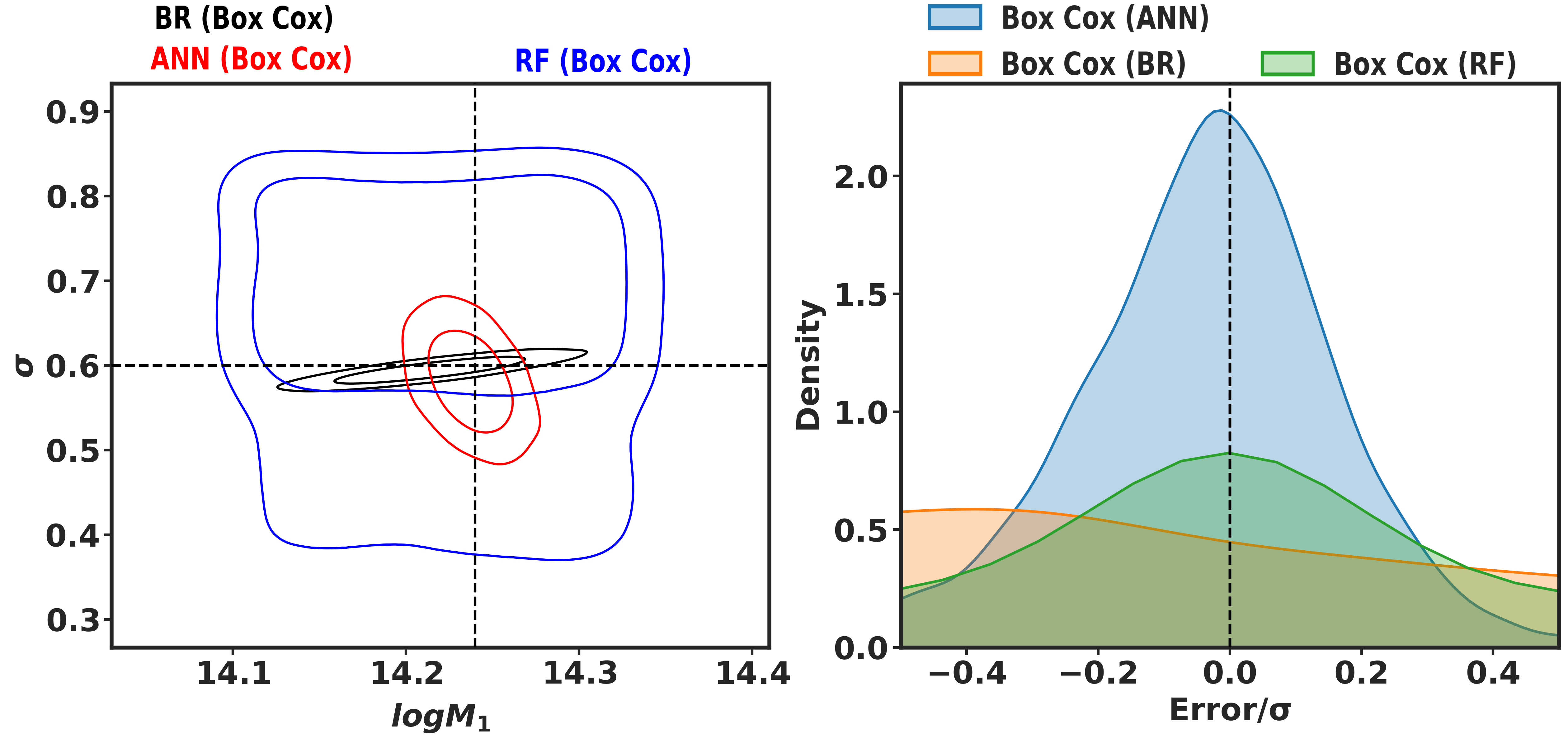}
    \caption{Exploring the impact of different ML methods within the small box. Left: Equipotential contours of the MCMC posterior from fitting the HOD model at $z=1.1$ using $w_\mathrm{p}$. The plot shows the convergence test between the RF, BR, and ANN models. Right: Density plot between the RF, BR, and ANN models as a function of Error/$\sigma$. The $\sigma$ is calculated from the jackknife covariance matrix.}
    \label{fig:comparison_ANN_BR_RF_smallbox}
\end{figure*}

\subsection{Algorithm Choice}
\label{sec:model_selection}

Fig.~\ref{fig:comparison_ANN_BR_RF_smallbox} shows the error distribution and the
HOD parameter likelihood contours for all three ML algorithms considered in this
work. In all three cases, the BC transformation was applied to the data. The
right panel on Fig.~\ref{fig:comparison_ANN_BR_RF_smallbox} shows that the ANN-based method is the least biased. The BR and RF methods
result in significantly wider likelihood contours compared to the ANN, which is the one closest to the truth. In the case of RF, the likelihood contours are completely dominated by the prior parameters. 

\subsection{Number of Samples}
\label{sec:effect_of_samplesize}
\begin{figure}
    \centering
	\includegraphics[width = \linewidth]{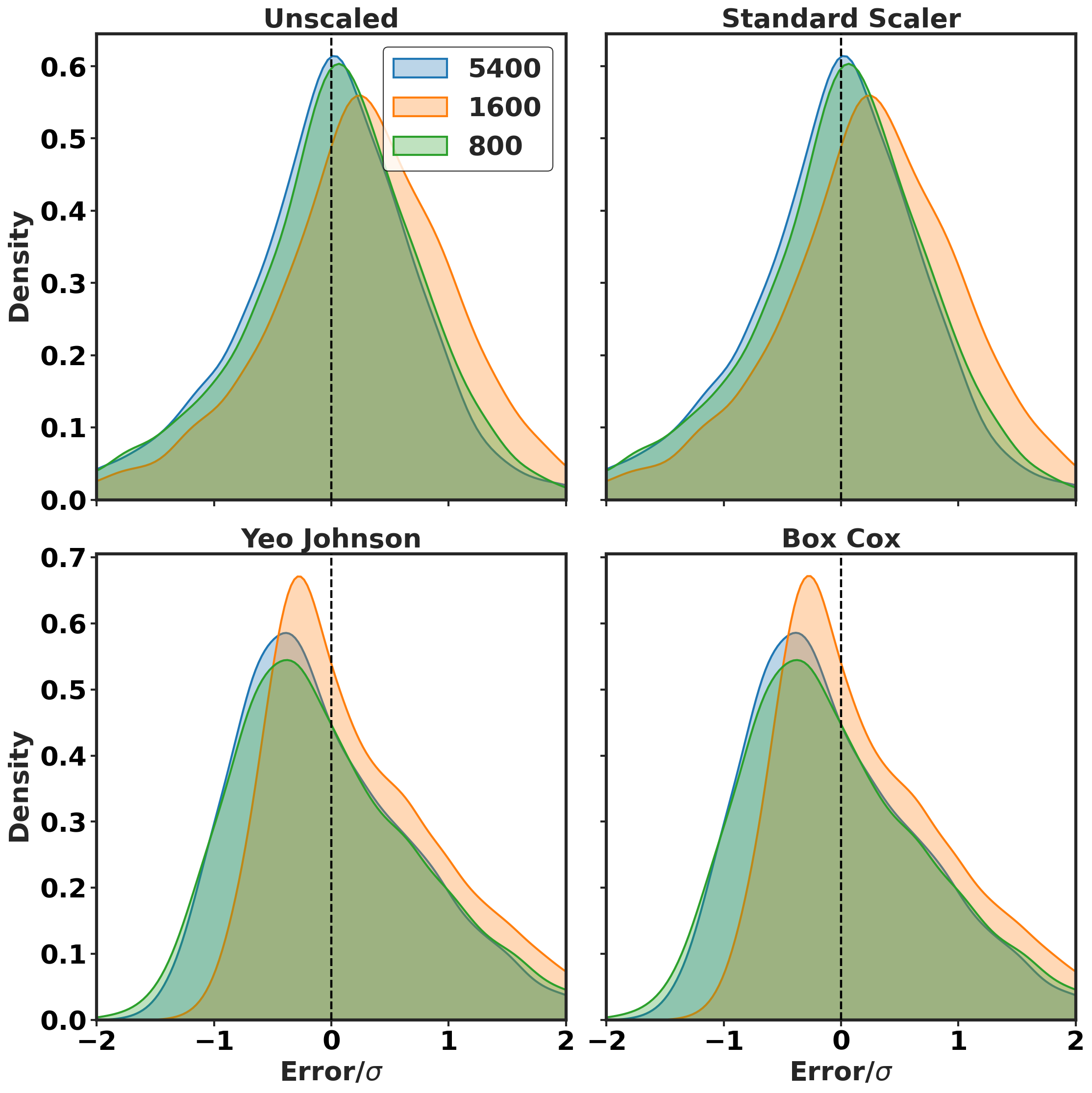}
    \caption{Effect of increasing data points on Bayesian Ridge Regressor for the small box. This plot compares the density plot between the training set of 800, 1600, and 5400 for unscaled, SC, YJ, and BC transformation as a function of Error/$\sigma$. The $\sigma$ is calculated from the jackknife covariance matrix.  }
    \label{fig:BR_density_small}
\end{figure}

\begin{figure}
    \centering
	\includegraphics[width =\linewidth]{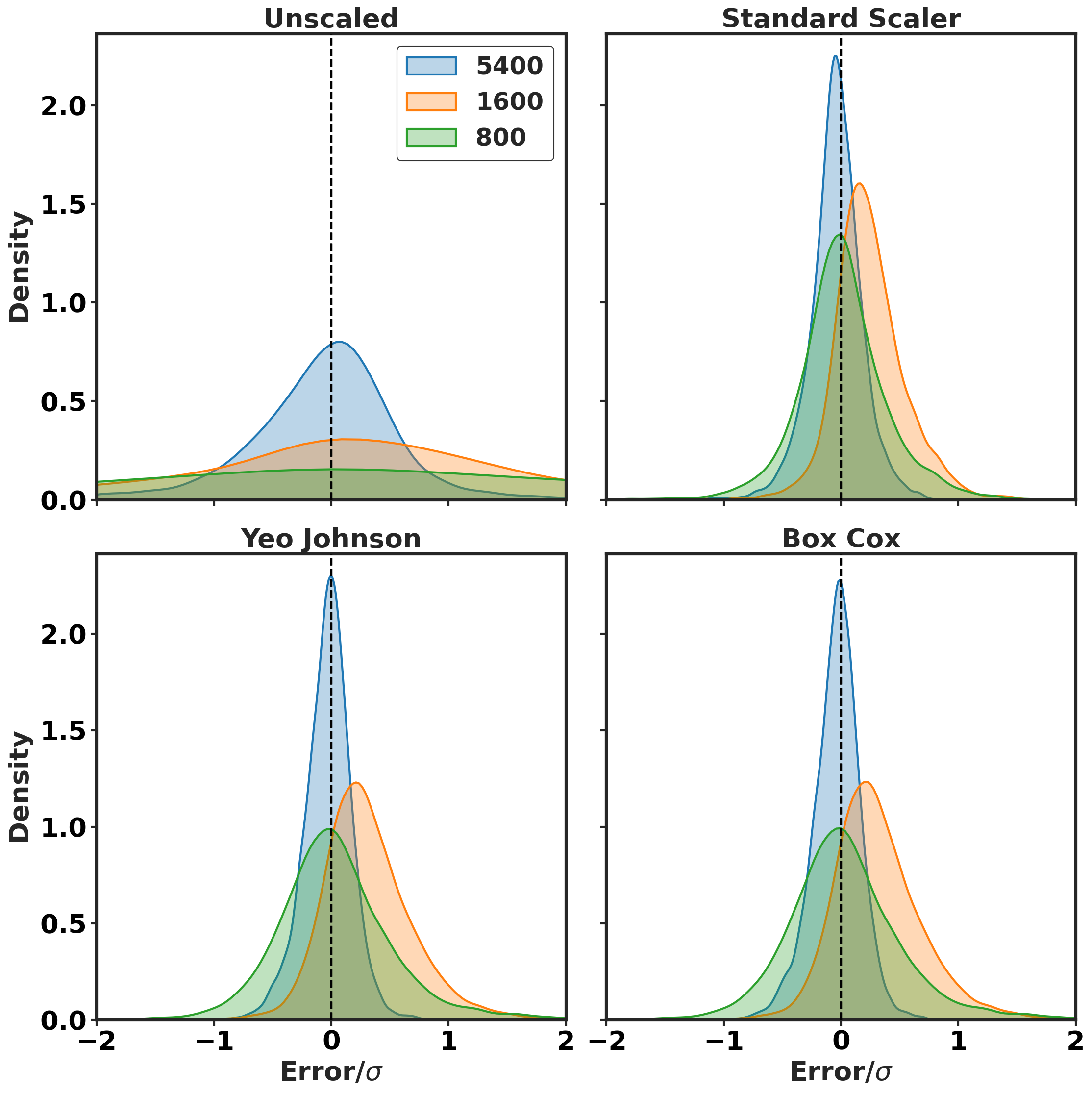}
    \caption{Effect of increasing data points on ANN for the small box. This plot compares the density plot between the training set of 800, 1600, and 5400 for unscaled, SC, YJ, and BC transformation as a function of Error/$\sigma$. The $\sigma$ is calculated from the jackknife covariance matrix. }
    \label{fig:ANN_density_small}
\end{figure}

We anticipate that all methods will eventually converge towards the truth as the size of the training sample increases. On the other hand, an excessively large training sample would defeat the primary purpose of the emulator, which is to obtain likelihoods with significantly less computational time.

Fig.~\ref{fig:BR_density_small}
shows the effect of increasing sample size on our results for the BR method. We
see that increasing the sample size by a factor of two and then a factor of
seven does not have a significant effect on the performance of the algorithm.

For the ANN algorithm, on the other hand, the improvement in the size of the
training data is clearly visible. The four panels in
Fig.~\ref{fig:ANN_density_small} shows the evolution of the error distribution
as the sample size first doubles and then increases by a factor of seven. The
improvement in this distribution, however, is not as drastic as one would hope.

These findings suggest that for the BR method, increasing the sample
size does not yield significant improvement, while for ANN, increasing the sample
size can improve performance.

\subsection{Prior Parameter Ranges}
\label{sec:prior}

\begin{figure*}
    \centering
	\includegraphics[width = \linewidth]{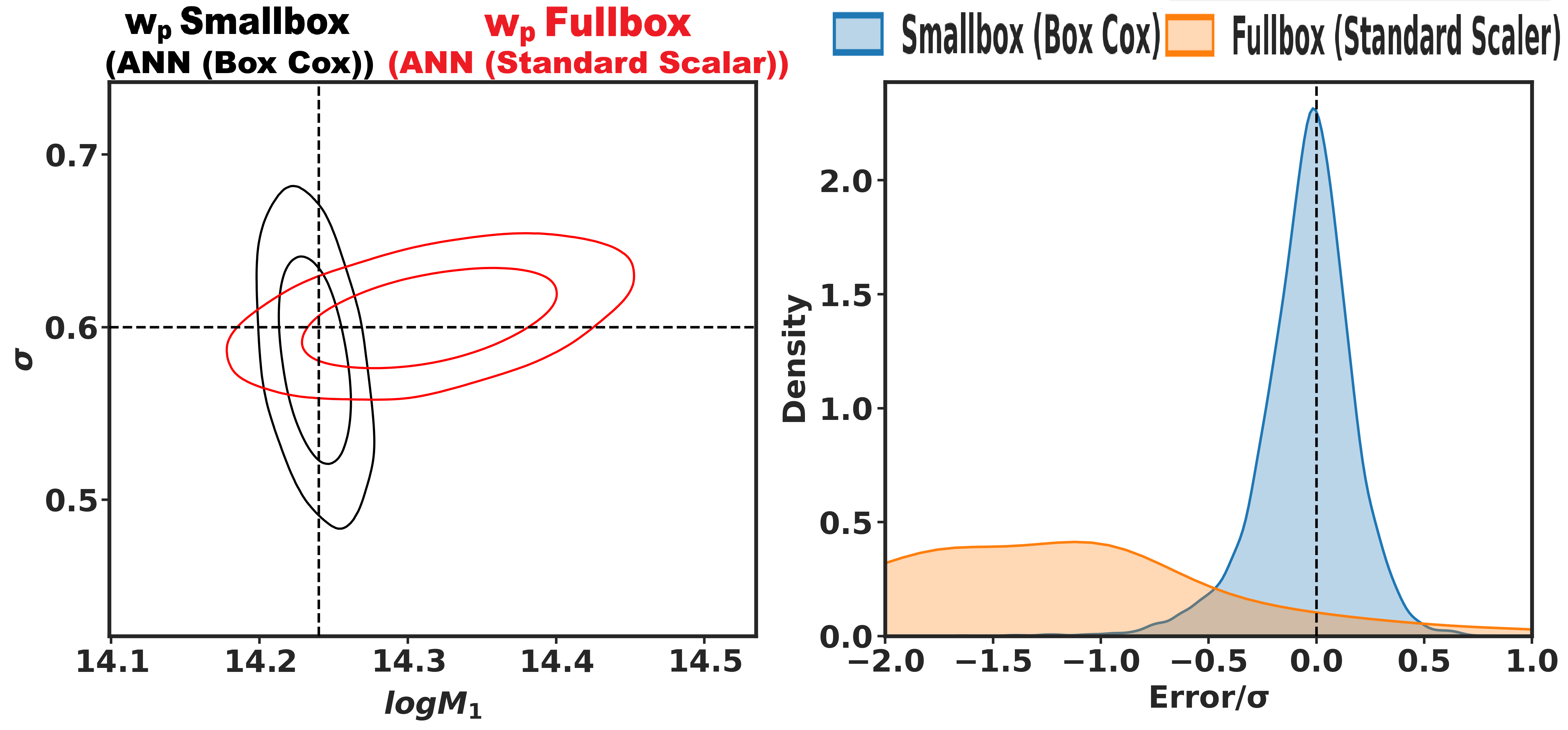}
    \caption{Error Comparison Between full Box and small box. Left: Equipotential contours of the MCMC posterior from fitting the HOD model at $z=1.1$ using $w_\mathrm{p}$. This plot shows the convergence between different box sizes. Right: Density plot between small box and full box as a function of Error/$\sigma$. The $\sigma$ is calculated from the jackknife covariance matrix.}
    \label{fig:comparison_fullbox_smallbox}
\end{figure*}

Fig.~\ref{fig:comparison_fullbox_smallbox} shows the effect of increasing the
prior range of HOD parameter space that the ML algorithm has to deal with. The
black and red contours on the left panel correspond to the small box and
full box respectively as defined in Section~\ref{sec:hod}. The contours were
derived with the ANN algorithm using the BC data transformation for the small box and SC for the full box, the method that was performing the best on a smaller box as described in Section~\ref{sec:model_selection}. From Fig.~\ref{fig:comparison_fullbox_smallbox} it is clear that even the best-performing algorithm fails to accurately reproduce the likelihood surfaces when the parameter range gets too large. In the limit of a very large number of training samples, we still expect the ANN to reproduce the true likelihood faithfully. This would, however, require a very large training sample and the
computational cost of training the network will start approaching that of a
straightforward MCMC chain. 


\begin{table*}
\renewcommand{\arraystretch}{1.8} 

\centering
    
    \begin{tabular}{|p{0.2\textwidth}||>
    {\centering\arraybackslash\hspace{0pt}}p{0.09\textwidth}|>
    {\centering\arraybackslash\hspace{0pt}}p{0.09\textwidth}|>
    {\centering\arraybackslash\hspace{0pt}}p{0.09\textwidth}|>
    {\centering\arraybackslash\hspace{0pt}}p{0.09\textwidth}|>
    {\centering\arraybackslash\hspace{0pt}}p{0.1\textwidth}|}
    
    \hline
    \multicolumn{1}{|c||}{\multirow{3}{*}{Observables}} & \multicolumn{5}{c|}{1 $\sigma$ error} \\ \cline{2-6} 
     & \multicolumn{5}{c|}{Small Box} \\ \cline{2-6} 
     & \multicolumn{1}{c|}{log $M_\mathrm{cut}$} & \multicolumn{1}{c|}{log $M_1$} & \multicolumn{1}{c|}{$\sigma$} & \multicolumn{1}{c|}{$\kappa$} & $\alpha$ \\ \hline
    \multicolumn{1}{|c||}{$w_\mathrm{p}$} & \multicolumn{1}{c|}{$12.9988_{-0.0111}^{+0.011}$} & \multicolumn{1}{c|}{$14.2414_{-0.0103}^{+0.0099}$} & \multicolumn{1}{c|}{$0.5955_{-0.0138}^{+0.0136}$} & \multicolumn{1}{c|}{$0.9859_{-0.0226}^{+0.023}$} & $0.4005_{-0.0106}^{+0.0101}$ \\ \hline
    \multicolumn{1}{|c||}{$w_\mathrm{p}$ + $VPF$} & \multicolumn{1}{c|}{$12.9989_{-0.011}^{+0.0103}$} & \multicolumn{1}{c|}{$14.2409_{-0.0096}^{+0.0097}$} & \multicolumn{1}{c|}{$0.5962_{-0.013}^{+0.0125}$} & \multicolumn{1}{c|}{$0.9856_{-0.0214}^{+0.0206}$} & $0.4001_{-0.0091}^{+0.0088}$ \\ \hline
    \multicolumn{1}{|c||}{$w_\mathrm{p}$ + $\xi_0$} & \multicolumn{1}{c|}{$12.9984_{-0.007}^{+0.0069}$} & \multicolumn{1}{c|}{$14.2416_{-0.008}^{+0.008}$} & \multicolumn{1}{c|}{$0.5954_{-0.0121}^{+0.0121}$} & \multicolumn{1}{c|}{$0.9848_{-0.0191}^{+0.0194}$} & $0.4007_{-0.0086}^{+0.0089}$ \\ \hline
    \multicolumn{1}{|c||}{$w_\mathrm{p}$ + $\xi_0$ + $\xi_2$} & \multicolumn{1}{c|}{$12.9988_{-0.0066}^{+0.0064}$} & \multicolumn{1}{c|}{$14.2416_{-0.0073}^{+0.0075}$} & \multicolumn{1}{c|}{$0.5956_{-0.0117}^{+0.0116}$} & \multicolumn{1}{c|}{$0.985_{-0.0185}^{+0.0182}$} & $0.4001_{-0.0085}^{+0.0085}$ \\ \hline
    \multicolumn{1}{|c||}{$w_\mathrm{p}$ + $\xi_0$ + $\xi_2$ + $\xi_4$} & \multicolumn{1}{c|}{$12.9983_{-0.0059}^{+0.0058}$} & \multicolumn{1}{c|}{$14.2411_{-0.0073}^{+0.0069}$} & \multicolumn{1}{c|}{$0.5953_{-0.0113}^{+0.0113}$} & \multicolumn{1}{c|}{$0.9841_{-0.0178}^{+0.0179}$} & $0.3998_{-0.0081}^{+0.0081}$ \\ \hline
    \multicolumn{1}{|c||}{$\xi_0$} & \multicolumn{1}{c|}{$12.9975_{-0.0135}^{+0.0132}$} & \multicolumn{1}{c|}{$14.2426_{-0.0113}^{+0.0113}$} & \multicolumn{1}{c|}{$0.5941_{-0.0337}^{+0.0326}$} & \multicolumn{1}{c|}{$0.9825_{-0.0558}^{+0.0583}$} & $0.4005_{-0.0262}^{+0.0267}$ \\ \hline
    \multicolumn{1}{|c||}{$\xi_0$ + $\xi_2$} & \multicolumn{1}{c|}{$12.9986_{-0.0122}^{+0.0117}$} & \multicolumn{1}{c|}{$14.2418_{-0.0099}^{+0.0101}$} & \multicolumn{1}{c|}{$0.5965_{-0.0263}^{+0.0276}$} & \multicolumn{1}{c|}{$0.9769_{-0.051}^{+0.0508}$} & $0.3984_{-0.0222}^{+0.0224}$ \\ \hline
    \multicolumn{1}{|c||}{$\xi_0$ + $\xi_2$ + $\xi_4$} & \multicolumn{1}{c|}{$12.9986_{-0.0113}^{+0.0111}$} & \multicolumn{1}{c|}{$14.2419_{-0.0096}^{+0.009}$} & \multicolumn{1}{c|}{$0.5984_{-0.0273}^{+0.0267}$} & \multicolumn{1}{c|}{$0.9707_{-0.0424}^{+0.0433}$} & $0.396_{-0.0178}^{+0.0177}$ \\ \hline
    \multicolumn{1}{|c||}{$w_\mathrm{p}$ + $VPF$ + $\xi_0$ + $\xi_2$ + $\xi_4$} & \multicolumn{1}{c|}{$12.9984_{-0.0056}^{+0.0056}$} & \multicolumn{1}{c|}{$14.2409_{-0.0068}^{+0.0067}$} & \multicolumn{1}{c|}{$0.596_{-0.0105}^{+0.0104}$} & \multicolumn{1}{c|}{$0.9826_{-0.0166}^{+0.0158}$} & $0.3999_{-0.0074}^{+0.0072}$ \\ \hline
    
    \end{tabular}
    \caption{1 $\sigma$ error for the small box. The values are calculated by running the MCMC chain and using the ANN models for $w_\mathrm{p}$, VPF, $\xi_0$, $\xi_2$, and $\xi_4$.}
    \label{tab:small_box}
\end{table*}

\begin{figure*}
    \centering
	\includegraphics[width = \linewidth]{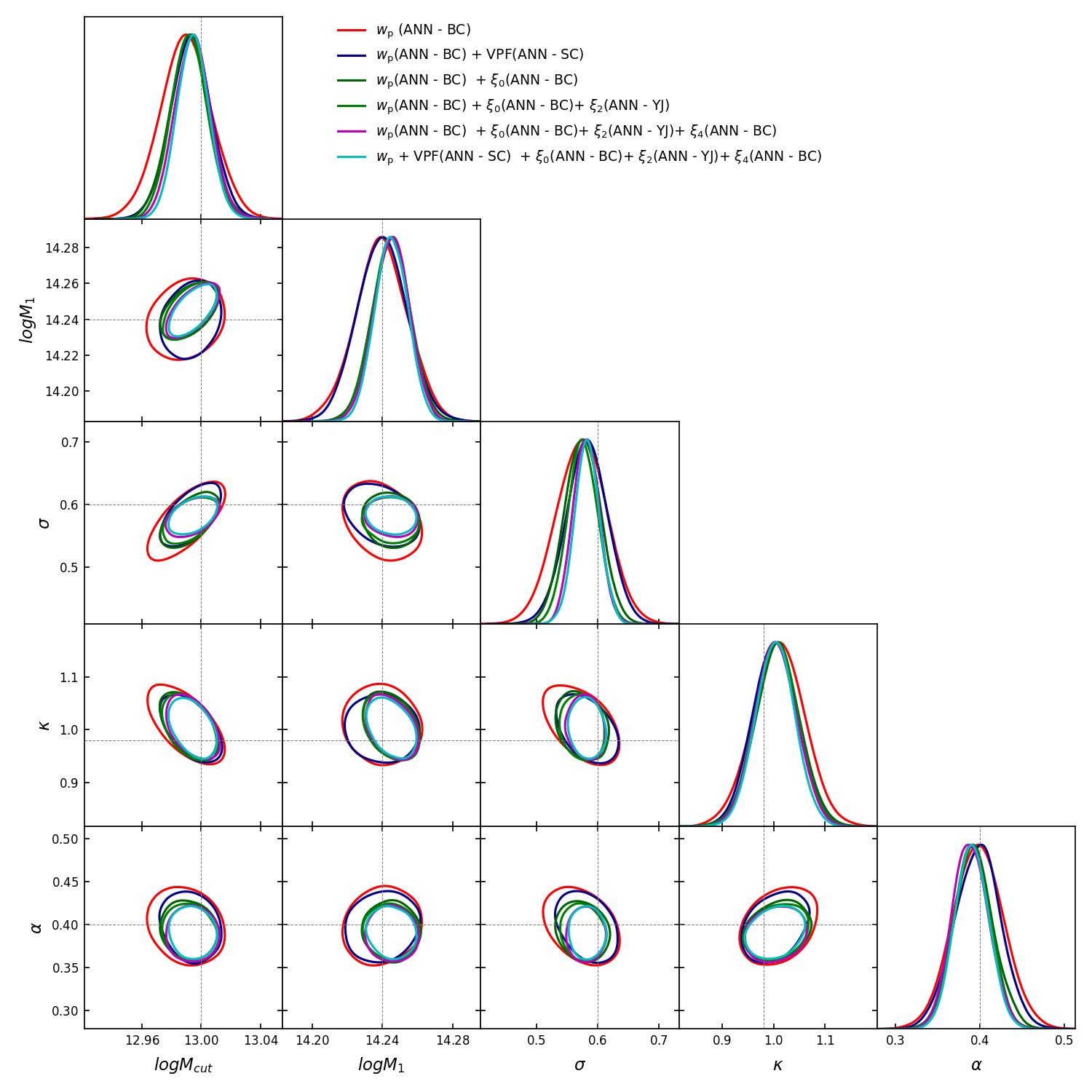}
    \caption{Equipotential contours of the MCMC posterior from fitting the HOD model at $z=1.1$. The ANN models are trained on the small box for $w_\mathrm{p}$, VPF, $\xi_0$, $\xi_2$, and $\xi_4$. The dashed lines denote the expected values of the HOD parameters. Contours are generated using the best-performing combinations of the ML models.}
    \label{fig:combined_ANN_small}
\end{figure*}

\begin{figure*}
    \centering
	\includegraphics[width = \linewidth]{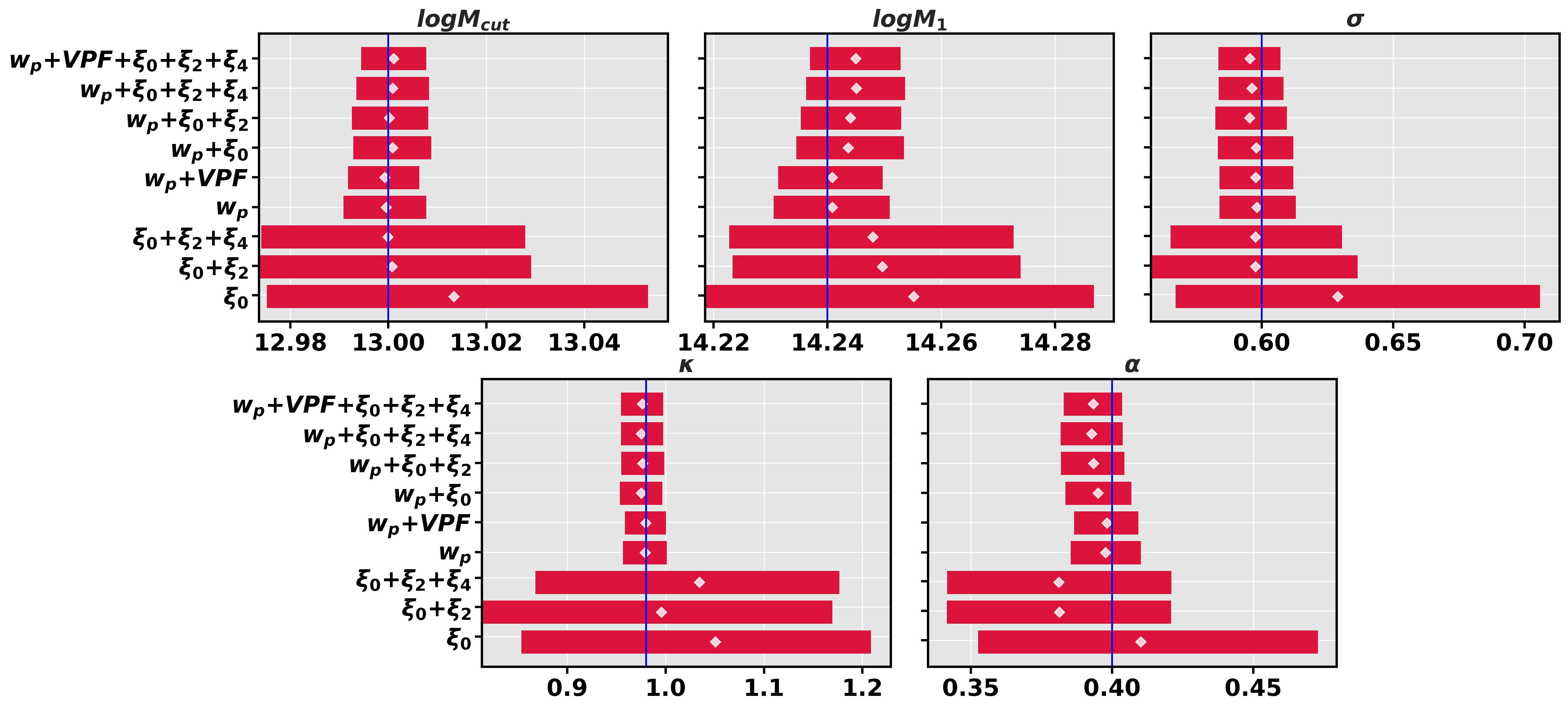}
    \caption{Effect of 1$\sigma$ constraint on HOD parameters for a different combination of statistics. The blue lines denote the expected values of the HOD parameters. The white marks at the center of the red bars represent the mean of the MCMC samples after discarding 200 steps as burn-in.}
    \label{fig:smallbox_ann_constraint_bar}
\end{figure*}


\section{Results and Conclusions}
\label{sec:conclusion}

Small-scale clustering of galaxies potentially contains useful information about cosmological parameters. To harness this information, however, one needs to first accurately model the nonlinear process of galaxy formation. For clustering measurements relevant to cosmology, the most important aspect is determining which types of galaxies occupy dark matter halos and exactly how they are distributed within those halos. The HOD-based family of models is one of the most widely used methods for modeling this galaxy-halo connection. Since generating HOD model predictions requires computationally expensive computations from N-body simulations, ML-based approaches that can significantly accelerate this computation have been used very frequently recently. 

We found that great care must be taken when modeling the dependence of small-scale clustering on HOD parameters. Transforming the input data, the choice of the ML approach, and the size of the HOD space under exploration, have a big impact on the robustness of the predictions. We found that very often the ML methods report acceptable formal performance scores, but when the inaccuracies in the ML modeling are propagated to the quantities that are actually important for cosmology, these small inaccuracies lead to unacceptable large biases. For example, the posterior likelihood in the HOD space can be strongly biased even when the ML pipeline reports a good performance score.

For the galaxy tracers and clustering measurements we used, we found that artificial neural networks (ANN) provided the best results among the models tested. Even for the ANN, the results were robust only when the HOD space was confined to the small box with a relatively narrow range of the parameters allowed and failed when we looked at larger parameter spaces. The selection of the small box range is largely driven by $M_\mathrm{cut}$, as this parameter is particularly sensitive to the filtering method we employed. In contrast, the other HOD parameters could potentially span larger ranges, as their distributions are less affected by this filtering process. The required restriction on the HOD parameter space was significant, being much smaller than the typical HOD parameter bounds allowed by currently available data. The ANN results become more robust with a larger set of training data as expected, but we found that the convergence to the true result with the training data size is not as fast as one would hope. The other two ML-based methods that we tried were the Bayesian Ridge and the Random Forest. These methods resulted in very large biases in the likelihood values even for the small box restriction.

Once the ML models achieve the desired robustness, the constraints on HOD
parameters can be obtained with very little computational time. We checked
how different combinations of $w_\mathrm{p}$, $VPF$, and $\xi_\ell$ 
enhanced the constraining power of small-scale clustering on the HOD parameter
space. The results of this exercise are shown in Table~\ref{tab:small_box} and
the 1$\sigma$ posterior contours are shown in Fig.~\ref{fig:combined_ANN_small}. The improvement in the HOD parameter constraints is shown in Fig.~\ref{fig:smallbox_ann_constraint_bar}. We find that the addition of VPF to $w_\mathrm{p}$ resulted in a slight improvement, but the combination of $w_\mathrm{p} + \xi_\mathrm{0}$ worked better than $w_\mathrm{p} + VPF$. These results are qualitatively in agreement with previous findings but want to warn the reader to treat the actual numbers (error bars, etc.) with caution. They should not be interpreted as actual forecasts for this type of analysis. The actual errors are very likely underestimated due to the usage of Jackknife covariance matrices and the prior effects from the small box restriction of allowable HOD space.

\section*{Acknowledgements}

This work is supported by DOE grants DE-SC0021165 and DE-SC0011840. 

This research used resources from the National Energy Research
Scientific Computing Center (NERSC), a U.S. Department of Energy Office of Science User Facility located at Lawrence Berkeley National Laboratory, operated under Contract No. DE-AC02-05CH11231.

AJ thanks Andrew Hearin and  Hanyu Zhang for useful discussions and comments.

\section*{Data Availability}
The data products related to this study will be shared upon request. The AbacusSummit simulations are publicly available to download from \url{https://abacusnbody.org/}.



\bibliographystyle{elsarticle-harv} 
\bibliography{reference}


\appendix

\section{Machine Learning Overview}
\label{sec:ML_overview}

We used supervised machine-learning techniques for regression to model the
dependence of small-scale clustering measurements on HOD parameters.
Specifically, we tested and compared three different regression algorithms -- ANN, RF, and BR regression.

\subsection{Artificial Neural Networks}

Artificial neural networks (ANNs) are computing systems inspired by biological
neural networks, designed to identify complex patterns in data. They consist of
an input layer, one or more hidden layers, and an output layer, each containing
neuron-like processing units connected by synaptic weights
\citep{2014arXiv1404.7828S}. By tuning these weights through backpropagation, ANNs
can approximate complex functions \citep{1986Natur.323..533R,
Hornik1991ApproximationCO}.

We implement ANNs using MLPRegressor in Scikit-Learn, a fully-connected
feedforward network. The HOD parameters were input features and clustering
measurements were the output. Using Hyperparameter tuning we determine the optimal
parameters like the number of layers/nodes, activation function, solver, and
regularization. Activation functions like "ReLU" and "tanh" were considered. To prevent overfitting, we used early stopping based on a validation set and dropout regularization \citep{Prechelt1998AutomaticES,Srivastava2014DropoutAS}.

\subsection{Random Forest Regression}

Random forest (RF) aggregates predictions from decision trees trained on randomized data subsets, which improves generalization and reduces overfitting~\citep{2001MachL..45....5B,2015arXiv151105741B}.RF regression averages predictions over all trees
\begin{equation}
\hat{y}_\mathrm{RF} = \frac{1}{B}\sum_{b=1}^B \hat{y}_\mathrm{b},
\end{equation}
where $\hat{y}_\mathrm{b}$ is the $b^{th}$ tree's prediction. We use
RandomForestRegressor in Scikit-Learn with hyperparameter tuning
\citep{2018arXiv180403515P}. It is known for its scalability, ability to handle high-dimensional data, and resistance to noise and outliers. While capable of handling large datasets, RF may become computationally expensive as the number of trees or features increases.

\subsection{Ridge Regression (BR)}

Ridge regression regularizes linear regression by penalizing the residual sum of
squares
\begin{equation}
L(\beta) = \sum_{i=1}^{n}{(y_i - \mathbf{x}_i\beta)^2} + \lambda|\beta|^2.
\end{equation}
The $\lambda$ hyperparameter controls regularization strength. This shrinks
coefficients to prevent overfitting. We used
BayesianRidge in Scikit-Learn, which estimates the regularization hyperparameter
($\lambda$) in a Bayesian framework \citep{MacKay1992BayesianI}. Ridge regression is similar to LASSO but uses L2 regularization.

Ridge regression is commonly fitted using coordinate descent optimization
algorithms. This involves iteratively
minimizing the objective function along each coordinate direction while holding
other coefficients fixed. The coordinate direction that leads to the largest
improvement in the objective is selected in each iteration. This continues until
convergence when changes along any coordinate cease to decrease the objective
further.

\section{Evaluating Machine Learning Model}
\label{sec:accuracy_score}

We evaluated our machine learning regression models using several numerical
measures that quantify the difference between predicted ($\hat{y}_i$) and actual
($y_i$) values. These accuracy metrics provided insight into model performance
and aided in selecting the best approach.

\subsection{Mean Squared Error}
The mean squared error (MSE) is computed by taking the average of the squared
differences between the predicted and true values

\begin{equation}
MSE = \frac{1}{n}\sum_{i=1}^{n}{(\hat{y}_i - y_i)^2},
\end{equation}
where $n$ is the number of samples.

Squaring the errors has the effect of heavily penalizing large deviations
compared to smaller ones. However, this disproportionate penalty also makes MSE
very sensitive to outliers that have large errors \citep{2005ClRes..30...79W}. MSE ranges from 0 to $\infty$, with lower values
indicating better model fit.

\subsection{Mean Absolute Error}
The mean absolute error (MAE) calculates the average magnitude of errors without
squaring
\begin{equation}
MAE = \frac{1}{n}\sum_{i=1}^{n}{|\hat{y}_i - y_i|}.
\end{equation}
Taking the absolute value of the differences avoids excessive penalties for
large errors. This makes MAE more robust to outliers compared to MSE
\citep{2001ApJ...549....1G,gmd-7-1247-2014}. MAE ranges from 0 to $\infty$, with 0 being a perfect
fit.

\subsection{R-squared}
The R-squared or coefficient of determination is
\begin{equation}
R^2 = 1 - \frac{RSS}{TSS},
\end{equation}
where RSS is the residual sum of squares
\begin{equation}
RSS = \sum_{i=1}^{n}{(y_i - \hat{y}_i)^2},
\end{equation}
and TSS is the total sum of squares
\begin{equation}
TSS = \sum_{i=1}^{n}{(y_i - \bar{y})^2}.
\end{equation}
Here for $i^{th}$ observation, $y_i$ is the actual value, $\hat{y}_i$ is the predicted value and $\bar{y}$ represents the mean of the actual values. $R^2$ compares the accuracy
of the model's predictions to the accuracy of simply predicting the mean in all
cases. It indicates the explanatory power of the model. $R^2$
values range from 0 to 1, with higher values indicating more variance explained
by the model.

\subsection{Adjusted R-Squared}

The R-squared metric can be adjusted to account for the number of predictors in
the model. The adjusted R-squared is defined as
\begin{equation}
R_{adj}^2 = 1 - (1-R^2)\frac{n-1}{n-p-1}
\end{equation}S
where $n$ is the sample size and $p$ is the number of explanatory variables in
the model.

Unlike regular $R^2$, the adjusted $R^2$ will penalize models for having
too many unnecessary predictors. It will always be less than or equal to the
unadjusted $R^2$. Generally, adding predictors improves the basic $R^2$
but can worsen the adjusted $R^2$ if they are redundant. This helps prevent
overfitting the training data \citep{theodoridis2015machine}.

For comparing regression models, the adjusted $R^2$ provides a more
conservative estimate that accounts for model complexity. However, it has
limitations when the sample size is small or the number of predictors is large
relative to the observations.

\begin{figure*}
	\includegraphics[width = \linewidth]{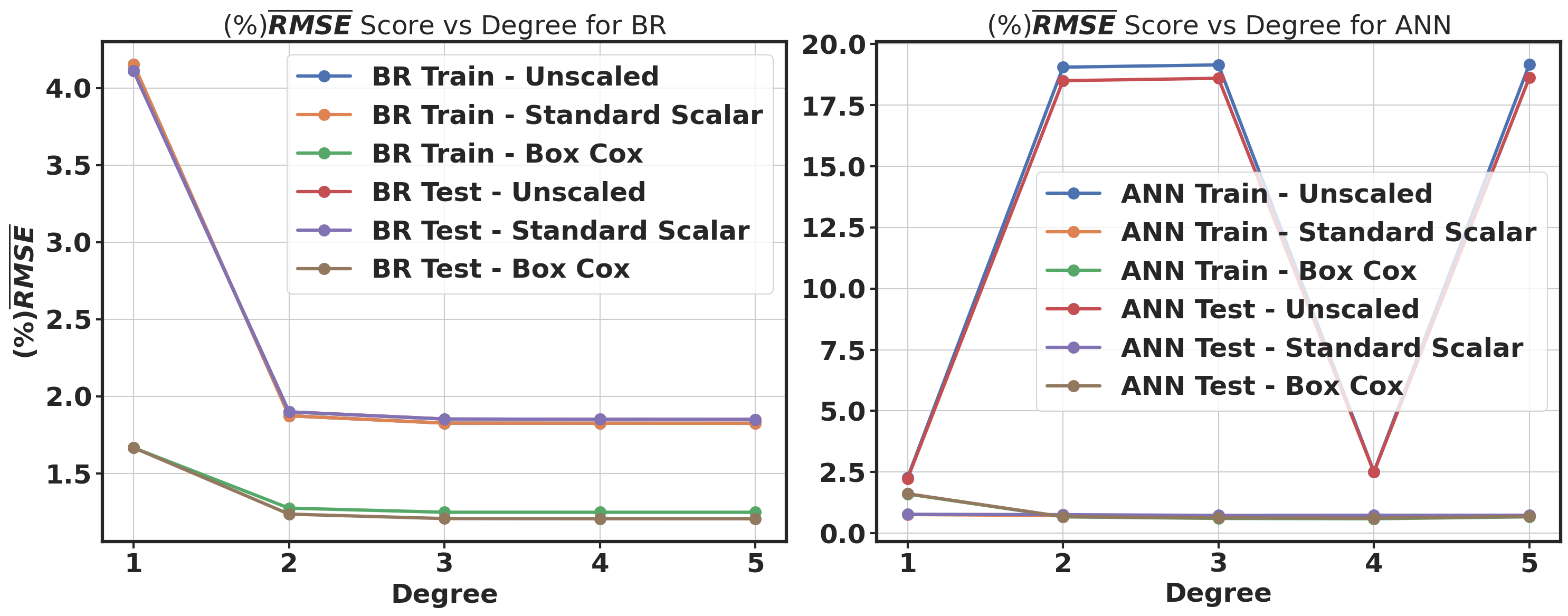}
    \caption{Comparison of average percent RMSE Scores between unscaled, SC, BC, and YJ on BR and ANN as a function of polynomial features on $w_\mathrm{p}$ for the small box train and test set. We calculate the average percent RMSE(see Section~\ref{sec:accuracy_test} for definition) as a function of degree polynomial. The left plot shows the average percent RMSE of the BR model. The left plot shows the average percent RMSE of the ANN model.}
    \label{fig:comparison_wp}
\end{figure*}

\begin{figure*}
	\includegraphics[width = \linewidth]{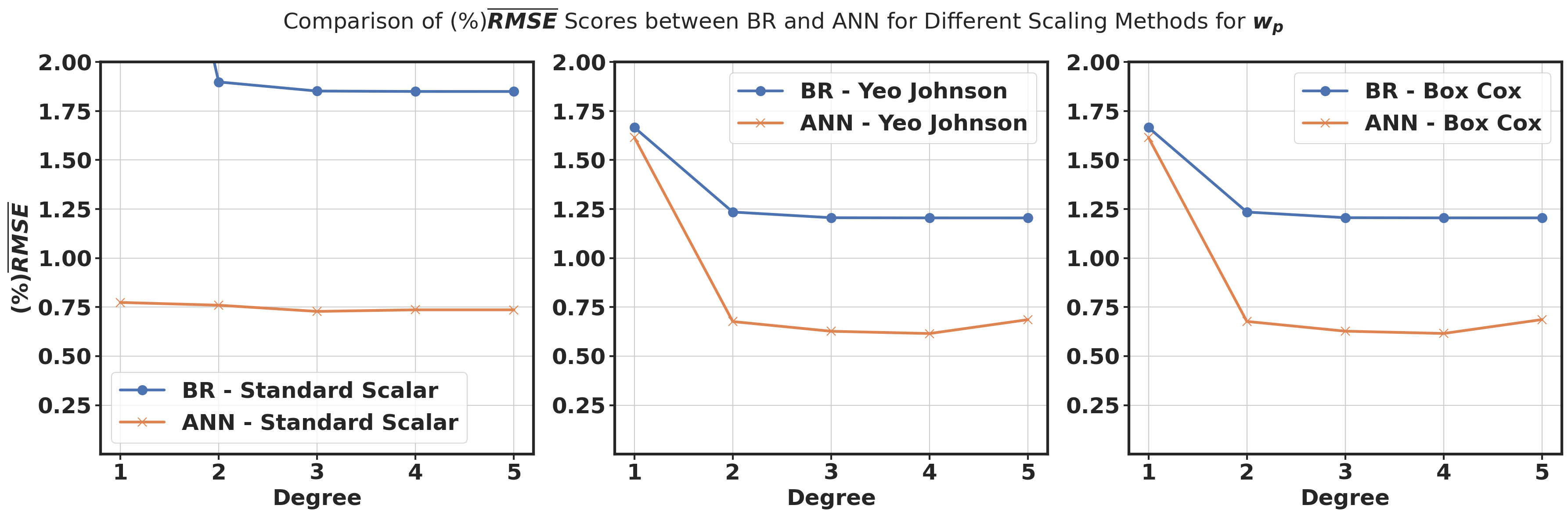}
    \caption{Comparison of average percent RMSE Scores between BR and ANN for SC, YJ, and BC on $w_\mathrm{p}$ test set. The left, middle, and right plot shows the average percent RMSE of SC, YJ, and BC transformations respectively.}
    \label{fig:rmse_wp}
\end{figure*}

\subsection{Choice of Polynomial Features}
Using polynomial features we can increase the number of features, capturing non-linear relationships in the data. Although a flexible model like ANN requires minimal preprocessing, incorporating polynomial features can reduce model complexity, leading to faster convergence. The selection of the polynomial degree is pivotal, as escalating it may result in model overfitting.

Fig.~\ref{fig:comparison_wp} shows the relationship between (\%)$\overline{RMSE}$ and the degree of polynomial on BR and ANN for $w_\mathrm{p}$ on the small box. (\%)$\overline{RMSE}$  score decreases with increasing the model complexity but decreases slowly after $2^{nd}$ order. We opt for a degree of 2, given marginal performance improvements beyond. Interestingly, ANN exhibits inferior performance with raw target data. 

Fig.~\ref{fig:rmse_wp} illustrates the comparison between ANN and BR. This finding aligns with Table~\ref{tab:multi_output_performance}, demonstrating that ANN outperforms BR at the same degree of polynomial.








\end{document}